\documentclass[fleqn,usenatbib]{mnras}

\usepackage{newtxtext,newtxmath}

\usepackage[T1]{fontenc}

\DeclareRobustCommand{\VAN}[3]{#2}
\let\VANthebibliography\thebibliography
\def\thebibliography{\DeclareRobustCommand{\VAN}[3]{##3}\VANthebibliography}

\usepackage{graphicx}	
\usepackage{amsmath}	
\usepackage{float}
\usepackage{colortbl}
\usepackage{arydshln}
\usepackage{array}
\usepackage{url}
\usepackage{natbib}
\usepackage{xcolor}
\usepackage{tikz}
\usepackage{isotope}
\definecolor{c1}{HTML}{44BB99}
\definecolor{c2}{HTML}{EEDD88}
\definecolor{c3}{HTML}{EE8866}

\title[Identification of external photoevaporation]{Line ratio identification of external photoevaporation}

\author[T. Peake et al.]{
Tyger Peake,$^{1}$\thanks{E-mail: t.peake@qmul.ac.uk}
Thomas J. Haworth,$^{1}$
Mari-Liis Aru$^{2}$
and William J. Henney$^{3}$
\\
$^{1}$Astronomy Unit, School of Physics and Astronomy, Queen Mary University of London, London E1 4NS, UK\\
$^{2}$European Southern Observatory, Karl-Schwarzschild-Strasse 2, D-85748 Garching bei München, Germany\\
$^{3}$Instituto de Radioastronomia y Astrofisica, Universidad Nacional Autonoma de Mexico, Apartado Postal 3-72, 58090 Morelia, Michoacan, Mexico
}

\date{Accepted XXX. Received YYY; in original form ZZZ}

\pubyear{\the\year{}}

\begin{document}
\label{firstpage}
\pagerange{\pageref{firstpage}--\pageref{lastpage}}
\maketitle

\begin{abstract}
External photoevaporation of protoplanetary discs, by massive O stars in stellar clusters, is thought to be a significant process in the evolution of a disc. It has been shown to result in significant mass loss and disc truncation, ultimately reducing the lifetime of the discs, and possibly affecting potential planet populations. It is a well-studied process in the Orion Nebula Cluster (ONC) where the cometary morphology of proplyds is spatially resolvable due to its proximity to Earth. However, we need to study external photoevaporation in additional stellar clusters to better understand its prevalence and significance more globally. Unfortunately, more massive stellar clusters where the majority of stars form are much farther away than the ONC. In these more distant clusters the proplyds are spatially unresolvable with current facilities, hence the cometary morphology is not a useful identification of external photoevaporation. Therefore, in order to identify and interpret external photoevaporation, the only observations we have are of spatially unresolved emission lines. To resolve this issue we have used the \textsc{cloudy} code to develop an approximate general model of the emission lines emanating from the hot ionized wind of a proplyd. We have used the model to determine which line ratios are most sensitive to the distance from an OB star, and found that the most sensitive line ratios vary by multiple orders of magnitude over an FUV field of between 10$^3$ G$_0$ to 10$^6$ G$_0$. By identifying spatial gradients of line ratios in stellar clusters, we can identify regions of ongoing external photoevaporation.

\end{abstract}




\section{Introduction}

Protoplanetary disks exist mainly around young stars \citep[e.g.][]{2001ApJ...553L.153H}. They are composed of gas and dust, providing the material for planet formation \citep[e.g.][]{2023ASPC..534..539M}. Therefore, it is likely that the dispersal of disk material will have an effect on planet formation and resultant planet populations \citep[e.g.][]{2012A&A...541A..97M, 2021A&A...656A..69E, 2023MNRAS.522.1939Q, 2025ApJ...979..120H}. Disc dispersal can be driven by internal processes: internal photoevaporation \citep[e.g.][]{2009ApJ...690.1539G,2012MNRAS.422.1880O, 2017RSOS....470114E, 2019MNRAS.487..691P, 2024A&A...690A.296S}, magnetohydrodynamic winds \citep[e.g.][]{1994ApJ...429..781S, 2019ApJ...874...90W, 2021ApJS..257...16B}, and  accretion onto the central star \citep[e.g.][]{1974MNRAS.168..603L, 2016A&A...591L...3M, 2021A&A...650A.196M}. However, most stars are born in stellar clusters \citep{2003ARA&A..41...57L} and disks may also evolve subject to external processes, such as stellar fly-bys \citep{2023EPJP..138...11C}, infall \citep{2023EPJP..138..272K, 2024A&A...683A.133G, 2024ApJ...972L...9W} and external irradiation (external ``photoevaporation'') by massive stars \citep{2022EPJP..137.1132W, 2025OJAp....8E..54A}.

The focus of this paper is on extending our ability to identify and interpret external photoevaporation. The first externally photoevaporating protoplanetary discs, located in the Orion Nebula Cluster (ONC), were observed by the Hubble Space Telescope  \citep[HST,][]{1993ApJ...410..696O,1994ApJ...436..194O,1998AJ....115..263O,2000AJ....119.2919B, 2008AJ....136.2136R}. They appear as disks with cometary (or teardrop) shaped ionization fronts directed towards the UV source responsible for their ionization. Since then, the ONC proplyds have been studied in great detail, including with KECK \citep{1999AJ....118.2350H}, JWST \citep{2022PASP..134e4301B, 2023arXiv231003552M, 2024Sci...383..988B}, VLT/MUSE \citep{2023A&A...673A.166K, 2023MNRAS.525.4129H, 2024A&A...687A..93A, 2024A&A...692A.137A} the VLA \citep{2023ApJ...954..127B} and ALMA \citep{2018ApJ...860...77E, 2020ApJ...894...74B, 2023ApJ...947....7B, 2025arXiv250302979B}. Theoretically, we expect the impact of external photoevaporation on the disks to be very significant. The high mass loss rates of gas and dust \citep{1998ApJ...499..758J, 2004ApJ...611..360A, 2016MNRAS.457.3593F, 2018MNRAS.481..452H, 2019MNRAS.485.3895H, 2023MNRAS.526.4315H, 2024A&A...681A..84G, 2025MNRAS.539.1414P} lead to rapid truncation, mass depletion and shorter lifetimes \citep[e.g.][]{2001MNRAS.325..449S, 2007MNRAS.376.1350C, 2019MNRAS.490.5678C,  2022MNRAS.514.2315C}. This in turn has been predicted to affect models of planet formation, for example due to limiting the mass reservoir for giant planet growth and migration \citep{2022MNRAS.515.4287W} and by restricting the flux of pebbles \citep{2023MNRAS.522.1939Q}. Other considerations of planet formation incude  \cite{2024A&A...689A.338H} and  \cite{2025ApJ...979..120H}. Overall we expect external photoevaporation can be very important for disk evolution and planet formation. However, star forming regions are complicated and evolving and we can only observe them at a snapshot in time, meaning the true impact can be difficult to interpret \citep{2019MNRAS.490.5478W, 2022MNRAS.512.3788Q, 2023MNRAS.520.5331W, 2025A&A...695A..74A}. Therefore, we require observations in regions other than the ONC. At similar distances to the ONC ($\sim400$\,pc) proplyds have also been discovered in the $<1\,$Myr cluster NGC 2024 \citep{2021MNRAS.501.3502H} and in the vicinity of the B star 42 Ori \citep{Kim2016, 2024ApJ...967..103B}. 

Despite the vast progress described above, in order to understand the true role external photoevaporation plays we need to be able to study its effect in even larger stellar clusters \citep{2008ApJ...675.1361F, 2020MNRAS.491..903W, 2022EPJP..137.1132W, 2025OJAp....8E..54A}, which are unfortunately more distant. For example, there are cases of more extreme clusters in Carina (e.g. Tr 14, Tr16) but existing facilities would be unable to spatially resolve ONC-like propylds at that distance of 2.3\,kpc. We hence require ways of identifying and interpreting external photoevaporation that do not rely on resolving a teardrop proplyd morphology. Even though 30m-class telescopes such as ESO's Extremely Large Telescope (ELT) \citep{2022JATIS...8b1510S, 2024arXiv240816396S, 2024SPIE13096E..12B} may eventually resolve proplyds at the distance of Carina, we would first be required to identify the best candidate externally photoevaporating disks from unresolved observations. This is particularly challenging given high UV environments are also associated with substantial nebular emission \citep[e.g.][]{2024A&A...685A.100I, 2024arXiv241205650R}. Therefore, we require models describing the expected emission properties of proplyds.

\cite{1998AJ....116..322H}  were the first to accurately model the brightness profile of H$\alpha$ in ONC proplyds. More recent models have incorporated emission lines of many more elements and constrain chemical abundances in specific proplyds, e.g. HST 10 and M42 \citep{2012MNRAS.426..614M,2013MNRAS.430.3406T}. Although these models accurately reproduce observed line intensities, they are computationally expensive and require often unknown elemental abundances. Other models have also considered the disk/PDR chemistry which has observational implications in the infrared and sub-mm for atomic and molecular lines \citep{2013ApJ...766L..23W,2023ApJ...947....7B, 2025ApJ...980..189G, 2025MNRAS.537..598K}, but our understanding of these processes is much less mature. We therefore aim to produce a simple yet general model of ionized gas emission lines, capable of spanning the desired parameter space of a proplyd. We use this model to test which line ratios are most capable of identifying ongoing external photoevaporation in stellar clusters when proplyds are not resolved.

The paper is organised as follows: in section \ref{Model} we describe the formation and implementation of the numerical model, in section \ref{sec:177-341 benchmark} we benchmark the model against the ONC proplyd 177-341W, in section \ref{results} we determine which line ratios are most sensitive to external irradiation and if spatial gradients in line ratios will be observable in stellar clusters, we follow this up with a discussion in section \ref{discussion}, and finally we summarise and conclude in section \ref{conclusion}.

\section{Model}\label{Model}

The purpose here is to develop a fast and simple model capable of predicting and explaining line ratio trends as a function of the UV field strength and properties of the proplyd. We simulate along the line of sight between the proplyd and the UV source. We assume that the ionized wind begins at the hydrogen ionization front and has a constant velocity equal to the local sound speed there. This results in a simple density profile, scaling with radius as $n \propto r^{-2}$. The proportionality constant depends on the proplyd's mass loss rate which is a function of: the properties of the proplyd, the UV source, and the distance between them. The density profile and UV source are then fed into the one-dimensional photoionization code \textsc{cloudy}, which determines: the hydrogen ionization front radius, ionization structure, temperature profile, and hence the radial emissivity profile of any given emission line. For the purpose of calculating observables we assume that the emissivity profile is a function of radius only and defined over a hemisphere (the cusp of the proplyd outwards). The emissivity profiles are then integrated to produce observables such as line ratios i.e. simulating the observation of a spatially unresolved proplyd. We run this model for populations of proplyds in order to determine which line ratios are most sensitive to the distance between the proplyd and the UV source, and whether spatial gradients in line ratios will be observable in stellar clusters. The following subsections elaborate on the model.

\subsection{Density profile in the ionized wind}

The physical model assumes a spherically diverging ionized wind initiated at the hydrogen ionization front. Applying conservation of mass to a spherically symmetric outflow, the density profile goes as
\begin{equation}\label{Density profile}
n = \frac{\dot{M}}{4 \pi r^2 v(r) \mu m_H}
\end{equation}
where $\dot{M}$ is the mass loss rate, $r$ is the radius, $v(r)$ is the radial velocity profile, $\mu$ is the mean molecular mass which we take as 0.6 for an \ion{H}{II} region, and $m_\text{H}$ is the hydrogen mass. For the reasons discussed below we assume that the velocity $v$ is radially constant and equal to the sound speed $c$ at the sonic point. A hydrogen I-front temperature of $T \approx 10^4$ K corresponds to a velocity of $v(r)=c \approx 10^6$ cm/s, which is the velocity that we use throughout. A more realistic velocity profile can be computed by simultaneously solving the mass and momentum equation, which we explore in appendix \ref{A1}. However, applying this model requires iteration of the ionization front radius $r_\text{IF}$ until $n_{\ion{H}{I}}(r_{\text{IF}})=n_{\ion{H}{II}}(r_{\text{IF}})$. In the constant velocity case the density profile is independent of the location of $r_{\text{IF}}$. Therefore, radiation transfer through the proplyd outflow naturally finds $r_{\text{IF}}$ and for efficiency we apply this case.

The mass loss rate $\dot{M}(r_\text{d},M_\text{h},G,\Sigma_{\text{au}})$ is determined by the \textsc{FRIED} grid of mass loss rates \citep{2023MNRAS.526.4315H} where: $r_\text{d}$ is the outer radius of the disc, $M_\text{h}$ the mass of the host star, $G$ the FUV radiation field found by integrating over the blackbody spectrum (corresponding to the UV source) between 6 eV and 13.6 eV, and $\Sigma_{\text{au}}$ is the surface density at 1\,au assuming that the surface density scales as $\Sigma(r) \propto r^{-1}$. For greater $r_\text{d}$ and $M_\text{h}$ the outer disc is less bound to the host star and the mass loss rate increases. For a stronger FUV field $G$ more energy is deposited into the disc, while for a greater disc mass $M_d \propto \Sigma_{\text{au}}$ more matter is available, both resulting in greater mass loss rates. Once $\dot{M}$ is specified by \textsc{FRIED} for a given proplyd the density profile is specified over the radial domain $r_\text{max}>r>r_{\text{IF}}$.  

The width of the ionization front can be approximated by the mean free path of an ionizing photon as $\Delta_{\text{IF}} \approx 1/\sigma_{\ion{H}{I}} n_{\ion{H}{I}}(r_{\text{IF}})$ where $\sigma_\text{H}$ is the hydrogen photoionization cross section and $n_{\text{IF}}$ is the hydrogen number density at the ionization front. The hydrogen photoionization cross section at $E_{\gamma}=13.6 \ \text{eV}$ is $\sigma_\text{H} \approx 6 \times 10^{-18} \ \text{cm}^{-2}$. We take a quarter of this value to account for the effects of radiation hardening \citep{2005ApJ...621..328H}. For typical proplyd outflows the \ion{H}{I} density at the I-front is $ \approx 10^4 \ \text{--} \ 10^6 \ \text{cm}^{-3}$ \citep[consistent with observations by][]{2023ApJ...954..127B, 2025arXiv250302979B}. The I-front width is then $\Delta_\text{IF} \approx 0.04 \ \text{--} \ 4 \ \textrm{au}$. Since we are interested in highly illuminated proplyds the density is on the upper end of this range and we comfortably satisfy $\Delta_{\text{IF}} \approx 0.04 \ \textrm{au} \ll r_{\text{IF}}$. Therefore, we will neglect the density structure across the ionization front, i.e. the transition between the ionized outflow and the photodissociation region (PDR).

In the shocked neutral shell of the PDR the matter is neutral and the density is approximately radially constant. For first-order models such as ours (where the structure across the ionization front is neglected) the density is discontinuous across the ionization front by a factor of $2c_{\text{ion}}^2/c_{\text{PDR}} ^2 = 2T_{\textrm{ion}}/T_{\textrm{PDR}} $, where the subscripts ion and PDR refer to the sound speed at either side of the ionization front. For example, assuming $T_{\textrm{PDR}} \approx 500$ K and $T_{\textrm{ion}} \approx 10^4$ K the discontinuity would be a factor of $\approx 40$. The large density discontinuity and small $\Delta_{\text{IF}}$ result in an extremely sharp fall in temperature and electron density. Consequently, the emissivities for both collisionally excited lines (CELs) and recombination lines (RLs) decline sharply in this region. Therefore, we can neglect their emissivities for $r<r_\text{IF}$. Once the simulation has reached $r_{\text{IF}}$ we end it.

\subsection{Radiation transfer with cloudy}

In order to determine the emission structure throughout the proplyds outflow we feed its density profile into the one dimensional photoionization code \textsc{cloudy} \textbf{\citep{2013RMxAA..49..137F, 2017RMxAA..53..385F, 2023RMxAA..59..327C}}. We use the preset Orion abundances since we make comparisons to ONC proplyds. We then specify the distance from the UV source $d$, the temperature of the star $T_*$, and its luminosity $L_*$. The distance is a proxy for the FUV field $G$, which for a given disc/star system sets the mass loss rate. Therefore, when referring to the strength of the UV environment we use $d$ and $G$ interchangeably. The code computes the spectrum incident on the proplyd and simulates radiation transfer through the outflow until our stopping point of $r=r_{\text{IF}}$. Proplyd photoionization flows are in the recombination-dominated regime \citep{2001ApJ...548..288L, 2001RMxAC..10...57H}. Therefore, an analytical estimate can be made for the ionization front radius by assuming that incident ionizing photons are balanced by hydrogen recombinations in the outflow. This results in the condition
\begin{equation}
    \frac{\dot{N}_{\text{EUV}}}{4\pi d^2} = \int^{\infty}_{r_{IF}} \alpha_\text{B} \ n_{\ion{H}{II}} \ n_e dr
\end{equation}
where in a simplified hydrogen only flow $n=2n_\text{e}=2n_{\ion{H}{II}}$, $\alpha_\text{B}=2.6 \times 10^{-13} \text{cm}^3/\text{s}$ is the case B hydrogen recombination coefficient \citep{1995MNRAS.272...41S}, $d$ is the distance from the ionizing source, and $\dot{N}_{\text{EUV}}$ is the rate of EUV photons produced by the UV source. Inserting the density profile from equation \ref{Density profile} and rearranging for $r_{\text{IF}}$ results in \citep{2022EPJP..137.1132W}
\begin{equation}\label{Ifront}
    r_{\text{IF}} = 1200 \left( \frac {\dot{M}}{10^{-8}\text{M}_\odot \text{yr}^{-1}} \right)^{2/3} \left( \frac{\dot{N}_{\text{EUV}}}{10^{45} \text{s}^{-1}} \right)^{-1/3}  \left( \frac{d}{\textrm{pc}} \right)^{2/3} \ \text{au}.
\end{equation}
%
The simulation finds the steady state ionization structure and temperature profile for $r>r_{\text{IF}}$. We are particularly interested in the radial emissivity, $j(r)$, of emission lines computed by \textsc{cloudy}. With this we can determine the observable properties of the proplyd. We include all species available in \textsc{cloudy} \citep{2017RMxAA..53..385F, 2023RMxAA..59..327C} and focus on the emission properties of a subset with high utility as described in section \ref{sensitive_line_ratios}.

\begin{figure}
    \centering
    \includegraphics[width=\columnwidth]{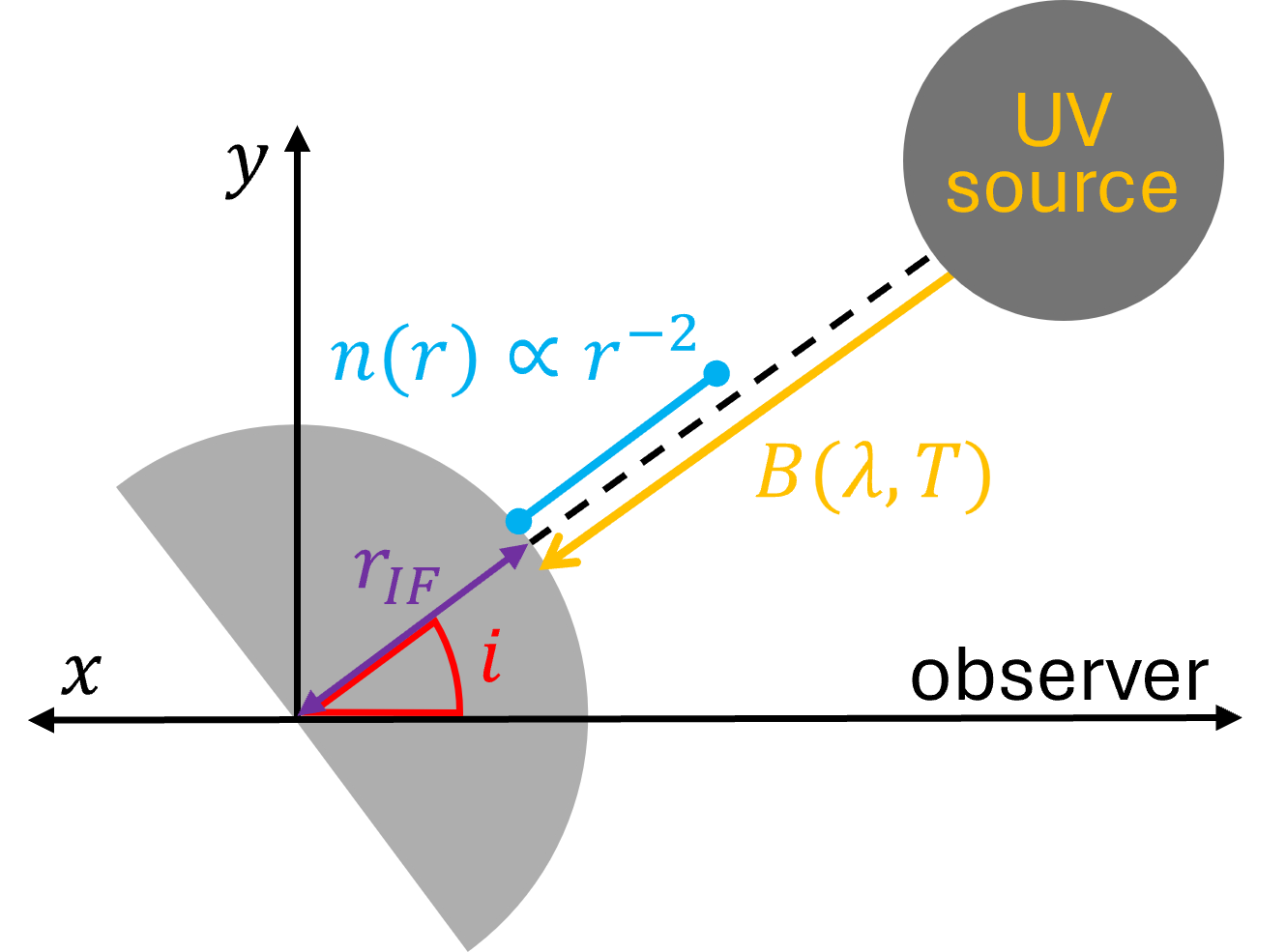}
    \caption{Schematic of the system. $(x,y)$ define a Cartesian coordinate grid and $r$ is the radial coordinate. The proplyd is the light gray hemisphere, with radius $r_{\text{IF}}$, facing the dark gray UV source. The UV source emits a blackbody spectrum $B(\lambda,T)$ which propagates down to the ionization front. The density profile is a function of radius only and is defined from $r_{\text{IF}}$ until the edge of the simulation domain (light blue line). The angle $i$ defines the orientation of the proplyd with respect to the observer.}
    \label{fig:schematic}
\end{figure}

\subsection{Proplyd observables}

In order to calculate observables we assume the emissivity, $j=j(r)$, is a function of radius only. It is defined over a hemisphere and is centred on the line of sight between the host star and the ionizing source. The geometry is shown in Figure \ref{fig:schematic}. The intensity of the line $I(y_\text{0})$, at a projected plane of sky distance $y_\text{0}$ from the proplyds centre, is found by numerical integration of
\begin{equation}\label{radial intensity}
    I(y_0)= \int_{-\infty}^{+\infty} j(x,y_0)e^{-\tau(x,y_0)}\textrm{d}x.
\end{equation}
The observer is located at $x=-\infty$ where $x$ is the coordinate along the observers line of sight, and $\tau(x,y_0)$ is the optical depth between $-\infty$ and $x$ along the line of sight through $y_\text{0}$ given by 
\begin{equation}
    \tau(x,y_0) = \int_{x'=-\infty}^{x'=x} d\tau(x',y_0)
\end{equation}
where $d\tau(x,y_0)$ is the optical depth through a cell length $dx$ along the line of sight defined by $y_0$.

We consider the optical depth in three distinct regions: the disc, the PDR, and the ionized wind. Firstly, we assume that the disc is opaque i.e. $d\tau(r\le r_\text{d}) \gg 1$. However, if $r_\text{d}^2 \ll r_\text{IF}^2$ the proportion of emission it absorbs is minimal and we assume that $d\tau(r \le r_\text{d})=0$. Secondly, for the PDR where $r_\text{d} < r_\text{i} \le r_{\text{IF}}$, we consider the two limiting cases: (i.) $d\tau=0$ everywhere, and (ii.) $d\tau \gg 1$ everywhere. Finally, for the ionized wind where $r_{\text{IF}}<r \le r_{\text{max}}$, we consider the radial optical depth $\tau_{\text{ion}}$ of each line throughout this region, calculated for us by \textsc{cloudy}. We neglect emission lines where $\tau_{\text{ion}} > 0.1$. We test these lines, assuming they're optically thin, and find that this does not result in the loss of any valuable line ratios. For lines where $\tau_{\text{ion}} < 0.1$ we include the line but neglect the optical depth in the wind due to its negligible effect, i.e. $d\tau(r>r_{\text{IF}})=0$. 

In the case that we want to simulate line ratios of unresolved proplyds we integrate the intensity of the line over the face of the proplyd, equation \ref{line_ratio}. In the limit $\textrm{d}\tau(r)=0$ for all r the integrals become just integrals of $j(r)$ over the volume of the proplyd. 
\begin{equation}\label{line_ratio}
\frac{l_1}{l_2}=\frac{2\pi\int_0^{\infty}I_1(y)y\textrm{d}y}{2\pi\int_0^{\infty}I_2(y)y\textrm{d}y} \xrightarrow{\textrm{d}\tau(r)=0 \ \forall \ r} \frac{4\pi\int_0^{\infty}j_1(r)r^2\textrm{d}r}{4\pi\int_0^{\infty}j_2(r)r^2\textrm{d}r}
\end{equation}
$l_1$ and $l_2$ are the power emitted from the proplyd from any two respective emission lines. We define a convergence criterion for the line luminosity integral as follows
\begin{equation}\label{luminosity convergence}
\int_0^{\frac{r_{\textrm{max}}}{2}}j_1(r)r^2\textrm{d}r \ \ge \ \frac{3}{4} \int_0^{r_{\textrm{max}}}j_1(r)r^2\textrm{d}r,
\end{equation}
and neglect emission lines that do not obey it, where $r_{\text{max}}$ is the extent of the radial domain in the simulation. This ensures that we only include lines which decay within the spatial domain so that the resultant line luminosity is not a function of $r_{\textrm{max}}.$
In the following sections, we will determine which lines have a strongly varying line ratio $\frac{l_1}{l_2}(d)$ as a function of distance $d$ from the UV source, and if we expect such trends to be observable in stellar clusters.

\section{Comparing with VLT/MUSE observations of ONC proplyds}
\label{sec:177-341 benchmark}

To ensure the validity of the model we compare it to observations of proplyds performed by the Multi-Unit Spectroscopic Explorer (MUSE) at the ESO Very Large Telescope (VLT) in narrow-field mode \citep[MUSE, ][]{2010SPIE.7735E..08B}.  The observations, collected under ESO programmes 104.C-0963(A) and 106.218X.001 (PI: Carlo F. Manara), cover the wavelength range $\sim$4750--9350 \AA\ with a field of view  of $\sim$7.5" $\times$ 7.5". The angular resolution of the images is measured to be $\text{FWHM} \approx 0.065\text{"}\text{-}0.080\text{"}$ at $\sim$6760 \AA. The observations are presented by \cite{2024A&A...687A..93A} for 12 proplyds in the ONC.  

\subsection{Benchmarking against proplyd 177-341W}
\label{sec:benchmarking}

\begin{figure}
    \centering
    \includegraphics[width=.8\columnwidth]{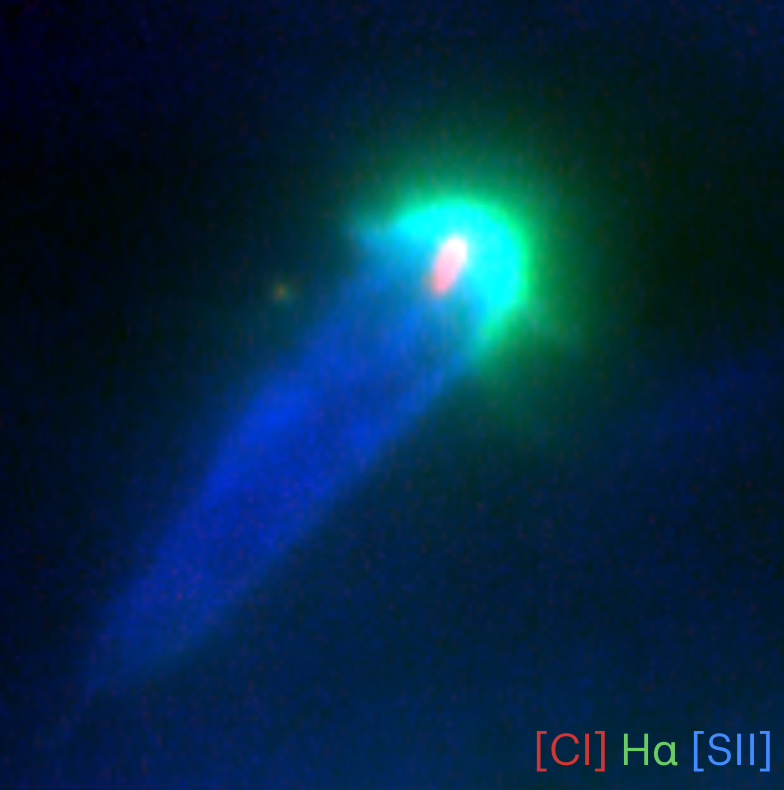}
    \caption{RGB image of proplyd 177-341W in the ONC, which combines [\ion{C}{I}], H$\alpha$, and [\ion{S}{II}] emission lines \citep{2024A&A...692A.137A}. We use this proplyd for comparison with our simplified model in section \ref{sec:benchmarking}. }
    \label{fig:proplyd-image}
\end{figure}

Firstly, we will focus on a detailed comparison against proplyd 177-341W. A three-colour image from VLT/MUSE observations of this proplyd is shown in Fig. \ref{fig:proplyd-image}. The properties of the proplyd and UV source system are shown in Table \ref{tab:177_341_properties}.

\begin{figure}
    \centering
    \includegraphics[width=\columnwidth]{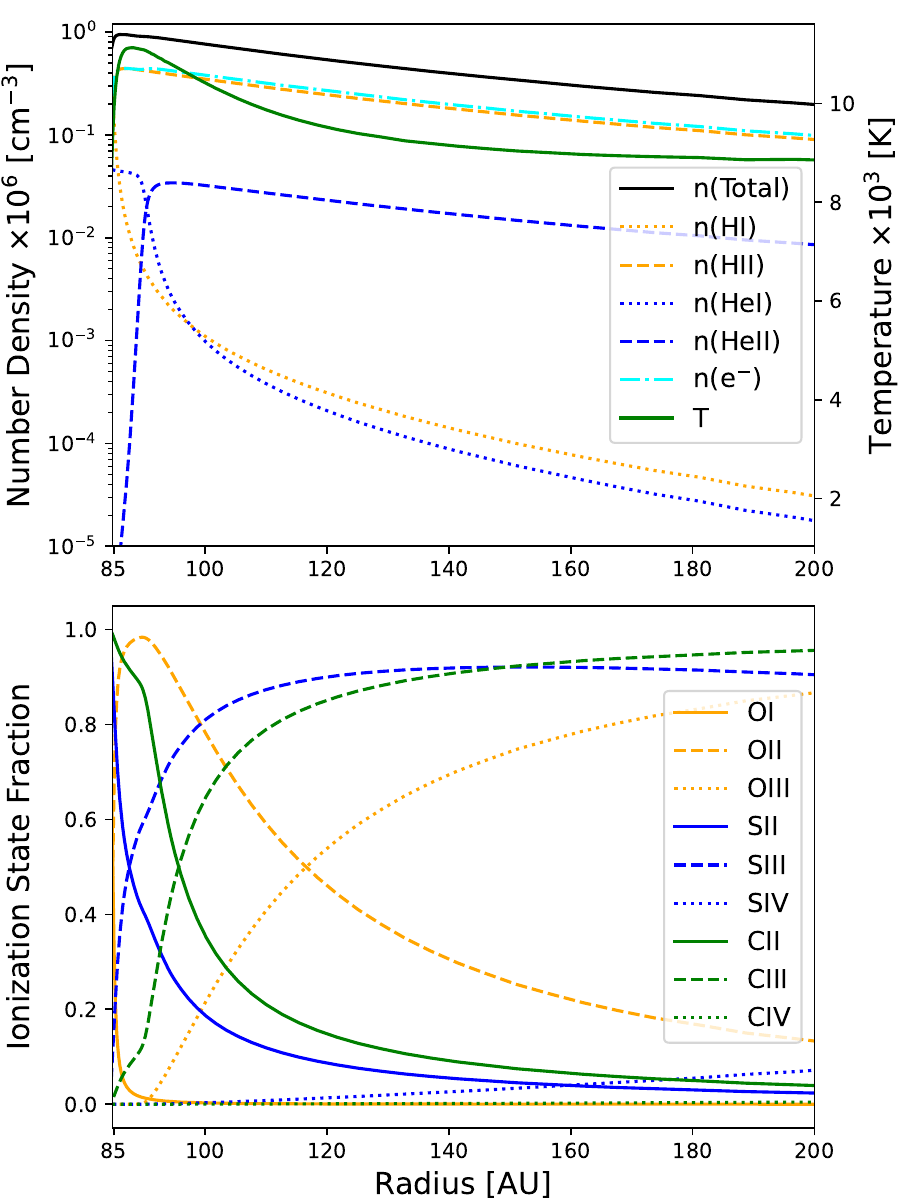}  
    \caption{Radial profiles of proplyd 177-341W. The upper plot shows the temperature, the total number density, and the number density of the most abundant species: $e^{-}$, \ion{H}{I}, \ion{H}{II}, \ion{He}{I}, \ion{He}{II}. The bottom plot gives the fraction of O, C, and S, in each ionization state.}
    \label{fig:177_341_profiles}
\end{figure}

{\renewcommand{\arraystretch}{1.5}
\begin{table}
    \centering
    \begin{tabular}{lr}
        \hline
        Disc Radius & $r_\text{d}=30.5 \ \text{au} \ ^{(1)}$ \\ 
        Mass of Host Star & $M_\text{h}=0.91 \ \text{M}_\odot \ ^{(2)}$ \\
        Projected Distance From UV Source & $d_{\perp}=0.049 \ \text{pc} \ ^{(2)}$ \\
        Peak H$\alpha$ Intensity Radius & $r_{\text{H} \alpha ,\text{obs}}=82 \ \text{au} \ ^{(2)}$ \\
        Mass Loss Rate & $\dot{M}_{\text{obs}}=5.11 \times 10^{-7} \ \text{M}_\odot/\text{yr} \ ^{(2)}$ \\
        Distance of ONC from Earth & $410 \ \text{pc} \  \ ^{(2)}$ \\
        UV Source Surface Temperature & $T_* = 39,000 \ \text{K} \ ^{(3)}$ \\
        UV Source Luminosity & $L_*=2.04 \times 10^5 \ \text{L}_\odot \ ^{(3)}$ \\
        Angle of Inclination & $i=80^{\circ} \ ^{(4)}$ \\
        Rate of Ionizing Photons & $\dot{N}_{\text{EUV}}=1 \times10^{49} \ \text{s}^{-1} \ \ \ \ \ $  \\
        FUV Field Strength at Proplyd & $G=8.5 \times10^{5} \ \text{G}_\text{0} \ \ \ \ \ $  \\
        Surface Density Normalization & $\Sigma_{\text{au}}=100 \ \text{g}/\text{cm}^{3} \ \ \ \ \ $ \\
        \hline
    \end{tabular}
    \caption{The properties of the proplyd 177-341W and the ionizing source $\Theta1$ Ori C from the following sources: (1) \protect\cite{2023ApJ...954..127B} (2) \protect\cite{2024A&A...687A..93A} (3) \protect\cite{2006A&A...448..351S} (4) \protect\cite{1999AJ....118.2350H}. $\dot{N}_{\text{EUV}}$ and $G$ are found by integrating over the blackbody spectrum, and we choose a representative value of $\Sigma_{\text{au}}$ due to uncertainty in the disc mass.} 
    \label{tab:177_341_properties}
\end{table}}
In the observations, rather than measuring $r_{\text{IF}}$, \cite{2024A&A...687A..93A} measure the radius $r_{\text{H}\alpha,\text{obs}}$, defined as the radius at which the intensity of $\text{H}\alpha$ emission is maximum. Since we want to make a comparison of the ionization structure it is imperative that this radius matches the equivalent radii in our simulation $r_{\text{H}\alpha,\text{sim}}$. The physical properties of 177-341W and the UV source system required for the simulation are given in Table \ref{tab:177_341_properties}. The angle of inclination has been measured as $i=80^{\circ}$ \citep{1999AJ....118.2350H}, so for simplicity we assume that $i=90^{\circ}$. The FRIED mass loss rate determined from these inputs, $\dot{M}_{\text{FRIED}}=2.0 \times 10^{-7} \ \text{M}_\odot/\text{yr}$, results in $r_{\text{H} \alpha,\text{sim}}  = 57 \ au  < r_{\text{H} \alpha ,\text{obs}}$. Therefore, for the purposes of this comparison $\dot{M}_{\text{FRIED}}$ was artificially scaled by a factor of 1.79 such that $r_{\text{H} \alpha ,\text{sim}}=r_{\text{H} \alpha ,\text{obs}}=82 \,\text{au}$, resulting in a simulation mass loss rate of $\dot{M}_{\text{sim}}=3.59\times10^{-7} \ \text{M}_\odot/\text{yr}$. This corresponds to a simulated ionization front radius of $r_{\text{IF}}=85 \ \text{au}$ (the analytical ionization front radius given by equation \ref{Ifront} is 75 au). We consider the order unity mismatch compared to the observed mass loss rate (see Table \ref{tab:177_341_properties}) a reasonably small error for a simple model designed to provide fast results. For the simulated proplyd, example radial profiles for density, temperature, and mixing ratio by ionization state are shown in Figure \ref{fig:177_341_profiles}.

\begin{figure*}
    \centering
    \includegraphics[width=\textwidth]{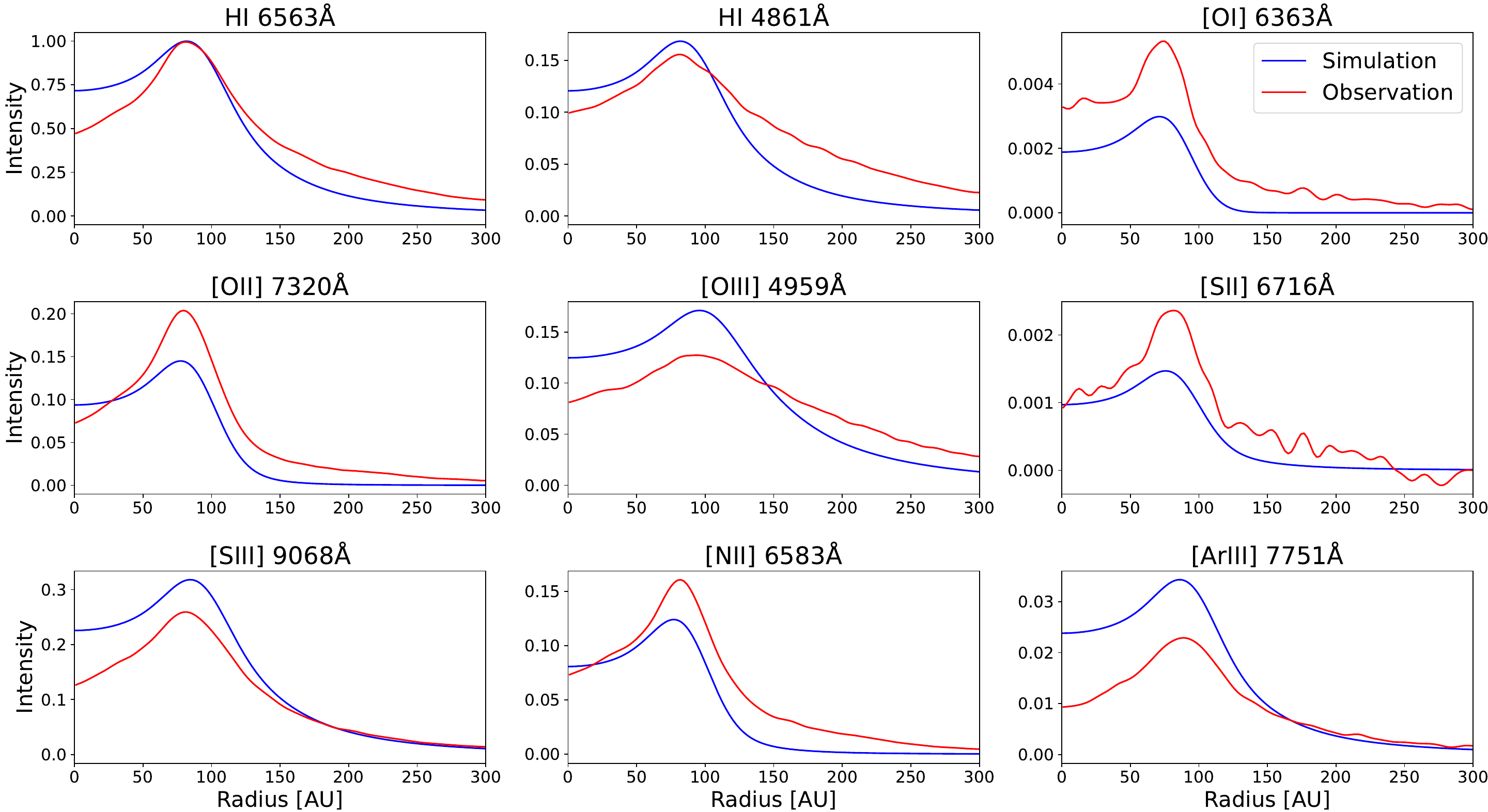} 
    \caption{Radial intensity profiles for proplyd 177-341W. The blue lines are from \textsc{cloudy} simulations, while the red are observations. We emphasize that the model has not been designed to accurately reproduce any single proplyd, but rather as a tool to predict general trends as a function of $G$. The mass loss rate from the simulated proplyd was scaled by a factor of 1.79 so that the radial location of the H$\alpha$ peaks are aligned. The simulations had extinction introduced so that the H$\beta$ simulation approximately matches the observation, and were then convolved by a point spread function. Finally, the simulations and observations were normalized so that their respective H$\alpha$ intensities were 1. 
    Across all lines the model is in good agreement with the observations given its simplicity.}
    \label{fig:177_341_comparison}
\end{figure*}

The observations and simulations of the radial intensity profiles, along the line of sight between the host star and UV source, are shown for 9 well observed emission lines in Figure \ref{fig:177_341_comparison}. In this case $r_\text{d}^2 \ll r_{\text{IF}}^2$, so we can neglect absorption by the disc, and we take the limit that $d\tau(\text{PDR})=0$.
The observations have not been extinction corrected (Aru et al. in prep.) so for this comparison, we introduce wavelength-dependent extinction between the source and observer into the model. The observed ratio of H$\alpha$ / H$\beta$ at the intensity peak $r_{\text{H}\alpha}$ is 6.1, and further out in the wind is $\sim4$, while the intrinsic value for recombination in the low density limit is 2.86 \citep{2006agna.book.....O}. The extinction may therefore either be locally higher near the I-front, or some other process (i.e. collisions) affects the line ratio.  For our purposes, where the goal is simply demonstrating that lines in the model have comparable intensities to the observations, we applied the CCM89 extinction law \citep{1989ApJ...345..245C} using the H$\alpha$ / H$\beta$ ratio at $r_{\text{H}\alpha}$. Assuming that $R_v=5.5$, appropriate for the Orion Trapezium, we applied the resulting wavelength dependent extinction law by reddening each line relative to H$\alpha$.
In addition, we account for the spatial resolution of the observation. We take a FWHM of $\approx0.0725\text{"} \approx 30 \ \text{au}$ (where we used that the ONC is $410$ pc from the Earth). The radial intensity profile is then convolved with a Gaussian with a standard deviation of $\sigma=\text{FWHM}/2\sqrt{2 \ \text{ln(2)}}$. 
Since we are only interested in the relative emission structure the simulated and observed intensity profiles have been scaled so that the H$\alpha$ lines match as closely as possible. The radial location of the peak in H$\alpha$ emission is already enforced by the imposition that $r_{\text{H}\alpha,\text{sim}}=r_{\text{H}\alpha,\text{obs}}$. Therefore, only the intensities must be scaled so that the H$\alpha$ intensities satisfy $I(r_{\text{H}\alpha,\text{sim}})=I(r_{\text{H}\alpha,\text{obs}})$. Finally, the intensity profiles were normalized by $I(r_{\text{H}\alpha,\text{sim}})$. As a result, the intensity units are dimensionless and none are stated in Figure \ref{fig:177_341_comparison}. Under the assumption that $i=90^\circ$, $\tau=0$, and $r^2=x^2+y_0^2$, we are plotting equation \ref{radial intensity} with the normalization 
\begin{equation}
    \frac{I(y_0)}{I(r_{\text{H}\alpha,\text{sim}})}=\frac{2}{I(r_{\text{H}\alpha,\text{sim}})}\int_{y_0}^\infty \frac{rj(r)\textrm{d}r}{\sqrt{r^2-{r_0}^2}}.
\end{equation}

In columns 2 and 3 of Table \ref{tab:177_341_results} the radius of the intensity maximum, which we measured in a consistent way for the observations and simulation, is shown for the 9 emission lines. Across all cases the positions of the peaks differ by no more than 3\,au. In particular, the larger peaks are very closely aligned, suggesting that the extended ionization structure is behaving as expected, e.g. [\ion{O}{III}] 4959 \AA, \ [\ion{Ar}{III}] 7751 \AA. The intensity of the peaks differ by no more than a factor of 2. This is a small discrepancy, given that the peaks of the observed intensities span a factor of $10^3$, only generic ONC abundances have been used, and we are not considering alternative sources of emission. In columns 5 and 6 of Table \ref{tab:177_341_results} the line ratio of H$\alpha$ against each line is shown for the observations and simulation. The line ratios differ by no more than a factor of $\approx3$. Since we are investigating only general trends, the line ratios exactly matching is unimportant. In section \ref{results} we investigate how line ratios such as these vary with distance from the UV source and its spectrum. 

{
\setlength{\tabcolsep}{12pt}
\renewcommand{\arraystretch}{1.5}

\begin{table*}
    \centering
    \begin{tabular}{lcccccc}
        \hline 
        Line & $r_{\text{obs}}(I_\text{max}) \ [\text{au}]$ & $r_{\text{sim}}(I_\text{max}) \ [\text{au}]$ & \% difference & $L_{\text{obs},\text{H}\alpha}/L_{\text{obs}}$ & $L_{\text{sim},\text{H}\alpha}/L_{\text{sim}}$ & \% difference \\
        \hline
        \text{\ion{H}{I} 6563 \AA} & 82 & 82 & 0.0 & 1.0 & 1.0 & 0.0 \\
        \text{\ion{H}{I} 4861 \AA} & 82 & 82 & 0.0 & 5.0 & 5.9 & 18 \\
        \text{[\ion{O}{I}] 6363 \AA} & 73 & 72 & 1.4 & 510 & 940 & 84 \\
        \text{[\ion{O}{II}] 7320 \AA} & 79 & 78 & 1.3 & 15 & 16 & 6.7 \\
        \text{[\ion{O}{III}] 4959 \AA} & 94 & 96 & 2.1 & 3.8 & 3.6 & 5.3 \\
        \text{[\ion{S}{II}] 6716 \AA} & 80 & 77 & 3.8 & 400 & 1300 & 230\\
        \text{[\ion{S}{III}] 9068 \AA} & 81 & 84 & 3.7 & 6.3 & 3.8 & 40 \\
        \text{[\ion{N}{II}] 6583 \AA} & 81 & 78 & 3.7 & 10 & 17 & 70\\
        \text{[\ion{Ar}{III}] 7751 \AA} & 88 & 87 & 1.1 & 52 & 33 & 37\\
        \hline
    \end{tabular}
    \caption{Properties of the intensity profiles shown in figure \ref{fig:177_341_comparison}, for both the simulations and observations. We show the radial position at which the intensity of each line peaks and the line ratio of H$\alpha$ to each line. For each result we show the percentage difference of the simulation with respect to the observation.}
    \label{tab:177_341_results}
\end{table*}
}

\subsection{Considering 2d models}

The simulated intensity profiles in Figure \ref{fig:177_341_comparison} are less symmetric around their peaks than their observational counterparts. For $r<r_{\text{IF}}$ the intensity falls off slower than for $r>r_{\text{IF}}$. In the 2d analytical model of \cite{1998ApJ...499..758J} the proplyd becomes an oblate spheroid, which has axial symmetry so we consider a plane through it, with a polar radius of $r_{\text{IF}}$ and an equatorial radius of $\approx1.3r_{\text{IF}}$. This model still assumes $n=n(r)\propto r^{-2}$. Therefore, since $j\propto n \ \text{--} \ n^2 \propto r^{-2} \ \text{--} \ r^{-4}$, the emissivity on the equator is $\approx 1.3^2 \ \text{--} \ 1.3^4 \approx 1.7 \ \text{--} \ 2.9$ times smaller than on the pole. The reduced emissivity would reduce the simulated intensity for $I(r<r_{\text{IF}})$, bringing them closer in line with the observations. Furthermore, the attenuation due to the PDR becomes more significant for $r<r_{\text{IF}}$. For both reasons, the intensity error in our models is greatest for smaller $r$. We ran a 2d model for 177-341W and found that, as expected, for $r<r_{IF}$ the intensity was reduced compared to the 1d model (for $r>r_{IF}$ the intensities were very similar). However, the contribution to the total luminosity from an annulus at $r$ scales as $r^{-1}$. Therefore, an annulus at radius $r<r_{\text{IF}}$ contributes less to the total luminosity, so its error is less significant. Given that each line is also impacted very similarly, this will have little affect on the final line ratio trends. 

While a 2d model is plausible, the increased computational expense is unnecessary and defeats the point of a fast model to predict general trends. It is clear from Figure \ref{fig:177_341_comparison} that the simulated intensity profiles are, for a simple model, reasonably similar to the observations. The difference in intensity for all lines across the whole radial domain is at worst a factor of $\approx 2$ times. Therefore, the 1d model with $n\propto r^{-2}$ proves to be sufficiently accurate and we will proceed with using it to predict general trends. 

\section{Results}\label{results}
\label{sec:results}

With a working model, we can determine which line ratios are most sensitive to the distance of the proplyd from the UV source. The most sensitive line ratios will be able to diagnose clusters in which external photoevaporation is ongoing. For every line under consideration we give line ratios as a function of $d$ (or FUV field) as supplementary data to this paper \footnote{\url{https://zenodo.org/records/15495918}}. In section \ref{sensitive_line_ratios} we show how line ratios vary as a function of distance from the UV source only, followed by a physical explanation in section \ref{Physical Explanation}. In section \ref{stellar_clusters} we consider the full parameter space of the proplyd/UV source system to determine whether spatial gradients in line ratios should be observable in stellar clusters, accounting for key observational effects.

\subsection{Predictions of most sensitive tracers of external irradiation}\label{sensitive_line_ratios}

For now we consider what we refer to as the base-case proplyd. This has a host star mass $M_\text{h}=0.7 \ \text{M}_\odot$, disc radius $r_\text{d}=50 \ \text{au},$ surface density normalization $\Sigma_{\text{au}}=100 \ \text{g}/\text{cm}^{3}$, $i=90^\circ$ and the stellar properties of $\Theta1$ Ori C given in Table \ref{tab:177_341_properties}. For this case we assume that $d\tau(r)=0$ for all $r$. We run models every $0.02$ pc from $d=0.05$ pc to $1.41$ pc corresponding to an FUV field of $10^6 \ \text{G}_0$ to $10^3 \ \text{G}_0$. At each $d$ we compute the luminosity of lines for all elements recorded by \textsc{CLOUDY} in the ionization states I, II, and III. Species which become the most abundant ionization state of their element at $r\leq r_{\text{IF}}$ are neglected, although we included H$\alpha$ as an exception. This is because we are not modelling the region $r \leq r_{\text{IF}}$ since we expect the lines that are bright in the ionized wind to be much weaker in the PDR. Restricting to wavelengths between 1000 \AA\ and 28.5 $\mu$m this leaves us with 6751 lines. 702 of these lines were neglected because their luminosities did not meet the convergence criterion. We checked the criterion for the $d=1.41$ pc simulation because this has the most extended ionization structure. 169 lines were neglected because their optical depth throughout the wind is greater than 0.1. For this we filtered using the $d=0.05$ pc simulation because this has the densest hydrogen ionization front. From the remaining 5911 lines we choose one emission line for each species. The intention is that this line is both bright and has line ratios which are sensitive to $d$. The line-selection procedure is given as follows.

\subsubsection{Line-selection procedure}

Firstly, we calculated the line ratio matrix $(l_\text{i}/l_\text{j})_\text{k}$, where k denotes the distance, and (i,j) denotes the combination of emission lines. For every pair of (i,j), we calculate the spatial mean $\overline{l}_{\text{ij}}$ and fit a straight line to $(l_\text{i}/l_\text{j})_\text{k}$ to obtain a gradient matrix $g_{\text{ij}}$. We then determine the Spearman's rank correlation coefficient matrix $R_{\text{ij}}$ in order to check the monotonicity of each line ratio as a function of distance from the UV source. Using these values, we define the line ratio measure $RM_{\text{ij}}$ as  
%
%
\begin{equation}\label{RM}
RM_{\text{ij}} =
\begin{cases}
\frac{\lvert  g_{\text{ij}} \rvert}{\overline{l}_{\text{ij}}} \times h\left(\frac{l_\text{i}}{l_{\text{H}\alpha}}\right) \times h\left(\frac{l_\text{j}}{l_{\text{H}\alpha}}\right), & \text{if} \ \ \ |R_{\text{ij}}| \geq 0.7 \\
0, & \text{if} \ \ \ |R_{\text{ij}}| < 0.7 \\
\end{cases}
\end{equation}
which is used to calculate the effectiveness of a given line ratio. $h(x)$ is a function defined as follows,
\begin{equation}\label{log_to_lin}
h(x) =
\begin{cases}
1+\log_{10}(x), & \text{if} \ \ \ x \ge 1 \\
\frac{1}{1-\log_{10}(x)}, & \text{if} \ \ \ x < 1 \\
\end{cases}
\end{equation}
and H$\alpha$ is used only as a reference line. Equation \ref{RM} was constructed so that the weighting from sensitivity, $\lvert  g_{\text{ij}}\rvert \ / \ \overline{l}_{\text{ij}}$, and luminosity $h(l_\text{i}/\text{H}\alpha)$ are on a comparable footing. The sensitivity spans a few orders of magnitude, but the luminosity varies by $\approx$ 50 so we consider the logged value of the luminosity given in equation \ref{log_to_lin}. In the case where $l_\text{i}>l_{\text{H}\alpha}$ then $h(l_\text{i}/l_{\text{H}\alpha})$ enhances $RM_{\text{ij}}$, whilst when $l_\text{i}<l_{\text{H}\alpha}$ it suppresses it. The ratio measure takes into account both the need for sensitive enough variation of $l_\text{i}/l_\text{j}$ with distance, and the need for sufficiently luminous lines for practical observations.
To find which lines are on average the most sensitive we define the line measure of a given line $x$ as 
\begin{equation}
    LM_\text{x} = \Sigma_\text{i} \ \ [ RM_{\text{xi}} \ + \ RM_{\text{ix}}].
\end{equation}
For each species the line with the largest $LM_\text{x}$ was chosen. This ensures that the lines most indicative of external photoevaporation are found without repeating lines of the same element and ionization state, which in most cases follow similar trends as each other. Exceptions to this arise when CELs of the same species and ionization state have emissivities which scale differently with density and temperature. We discuss examples of this in sections \ref{Density in the Outflow} and \ref{Temperature in the Outflow}.

\begin{figure}
    \centering
    \includegraphics[width=\columnwidth]{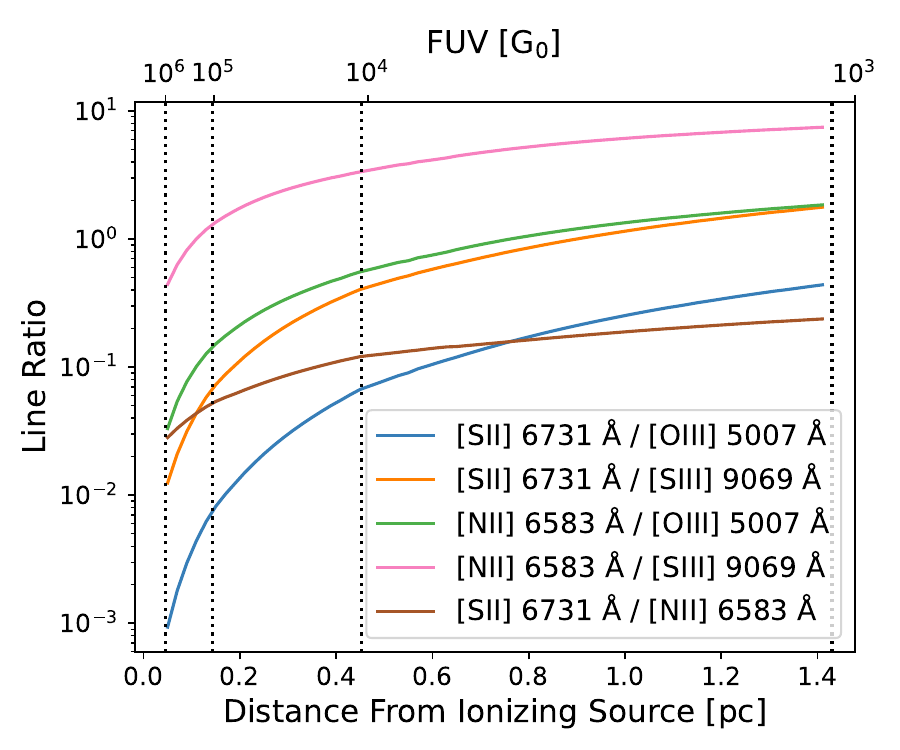}  
    \caption{Five example line ratios as a function of distance between our base-case proplyd and the UV source $\Theta1$ Ori C. We spanned $d=0.05$ pc to $d=1.41$ pc corresponding to FUV fields of between $10^3$ G$_\text{0}$ and $10^6$ G$_\text{0}$. 
    }
    \label{fig:Best_Line_Ratios}
\end{figure}

Considering the MUSE wavelength range, four of the earliest lines to appear in the line-selection procedure were [\ion{O}{III}] 5007 \AA, [\ion{S}{III}] 9069 \AA, [\ion{S}{II}] 6731 \AA, and [\ion{N}{II}] 6583 \AA. Between these lines the 5 most sensitive line ratios in order of sensitivity are: [\ion{S}{II}] 6731 \AA\ / [\ion{O}{III}] 5007 \AA, [\ion{S}{II}] 6731 \AA\ / [\ion{S}{III}] 9069 \AA, [\ion{N}{II}] 6583 \AA\ / [\ion{O}{III}] 5007 \AA, [\ion{N}{II}] 6583 \AA\ / [\ion{S}{III}] 9069 \AA, and [\ion{S}{II}] 6731 \AA\ / [\ion{N}{II}] 6583 \AA. They are shown as a function of distance (and FUV field) in Figure \ref{fig:Best_Line_Ratios}. The most sensitive line ratio [\ion{S}{II}] 6731 \AA\ / [\ion{O}{III}] 5007 \AA\ varies by a factor of 470 over an FUV field ranging from $10^3 \ \text{--} \ 10^6 \ \text{G}_\text{0}$. While, the weakest [\ion{N}{II}] 6583 \AA \ / [\ion{S}{II}] 6731 \AA \ still varies by a factor of 8.5. In section \ref{stellar_clusters} we consider, using these two line ratios as examples, whether spatial line ratio trends should be observable in stellar clusters.

\subsubsection{Sensitivity Tables}

The line-selection procedure was performed separately for both the range of wavelengths observable by MUSE, $4,800$ \AA\ -- $9,300$ \AA\, and the wavelength range, $1,150$ \AA\ -- 28.5 $\mu$m. A sensitivity Table between 20 of the selected emission lines with the largest $LM_x$ is given in Tables \ref{tab:line_ratios_MUSE} and \ref{tab:line_ratios_ALL} for the two wavelength ranges, respectively. We emphasize here that these are only a subset of all lines chosen based on $LM_x$ and that many more useful lines can be found in the full dataset through the link provided in the data availability section at the end of this paper. We define the factor by which a line ratio changes over the range $d=0.05$ pc to $d=1.41$ pc (or $10^3$ G$_\text{0}$ \text{--} $10^6$ G$_\text{0}$) as $f_{\text{ij}}$ (we will also use $f$ when referring to a specific element in the matrix). The values in the Table are then
\begin{equation}
\text{Sensitivity Table Values} =
\begin{cases}
\log_{10}(f_{\text{ij}}), & \text{if} \ \ \ |R_{\text{ij}}| \ge 0.7 \\
\text{-}, & \text{if} \ \ \ |R_{\text{ij}}| < 0.7 \\
\end{cases}
\end{equation}
where the signs simply correspond to positive if a line ratio increases with increasing distance, and negative otherwise. These values correspond to the factor by which the line ratios change in Figure \ref{fig:Best_Line_Ratios}. The highlighted colours correspond to 
\begin{equation}\label{colours}
\text{Colours} =
\begin{cases}
\text{\colorbox{c1}{Green}}, & \text{if} \ \ \ \lvert \log_{10}(f_{\text{ij}}) \rvert \ge 1 \\
\text{\colorbox{c2}{Yellow}}, & \text{if} \ \ \ 0.3 \le \lvert \log_{10}(f_{\text{ij}}) \vert < 1 \\
\text{\colorbox{c3}{Orange}}, & \text{if} \ \ \ \lvert \log_{10}(f_{\text{ij}}) \rvert < 0.3. \\
\end{cases}
\end{equation}
The indices i and j span the selected lines and denote the column and row, respectively. We use the metric $f_{\text{ij}}$ which does not consider the line luminosity because we have already filtered by luminosity, and are now only concerned with the sensitivity of the line with respect to $d$. Empty entries occur when $|R_{\text{ij}}| < 0.7$ in order to ensure a consistent trend across all distances. In both Tables the lines are ordered from left to right along the top row by descending value of $LM_\text{x}$. Therefore, the line ratios $l_\text{i}/l_{\text{j}}$ towards the upper left hand side are brighter and hence more feasibly observed. The first four selected lines, shown in Figure \ref{fig:Best_Line_Ratios}, populate the upper left hand corner of Table \ref{tab:line_ratios_MUSE}. In appendix \ref{A1} we determine that reproducing Table \ref{tab:line_ratios_MUSE} with a Parker wind density profile has little effect on the line ratio trends. Therefore, we can be confident in the validity of the simplified constant velocity wind.

{
\setlength{\tabcolsep}{2.5pt}
\renewcommand{\arraystretch}{1.5}
\begin{table*}
    \centering
    \begin{tabular}{c|ccccccccccccccccccccc}
        \hline
        species & \ion{O}{III} & \ion{H}{$\alpha$} & \ion{S}{III} & \ion{S}{II} & \ion{He}{I} & \ion{Ar}{III} & \ion{N}{II} & \ion{O}{II} & \ion{Cl}{III} & \ion{Fe}{III} & \ion{Fe}{II} & \ion{Ni}{II} & \ion{Si}{III} & \ion{C}{II} & \ion{Cr}{II} & \ion{Cl}{II} & \ion{Al}{II} & \ion{Mg}{II} & \ion{P}{II} & \ion{B}{II} \\ ($\lambda$ [\AA]) & \scriptsize{(5007)} & \scriptsize{(6563)} & \scriptsize{(9069)} & \scriptsize{(6731)} & \scriptsize{(7065)} & \scriptsize{(7136)} & \scriptsize{(6583)} & \scriptsize{(7320)} & \scriptsize{(8434)} & \scriptsize{(4986)} & \scriptsize{(7155)} & \scriptsize{(7378)} & \scriptsize{(8224)} & \scriptsize{(5113)} & \scriptsize{(8000)} & \scriptsize{(8579)} & \scriptsize{(7064)} & \scriptsize{(9218)} & \scriptsize{(7876)} & \scriptsize{(7030)} \\  \hline \ion{O}{III} \scriptsize{(5007)} & - & \cellcolor{c2} 0.77 & \cellcolor{c2} 0.52 & \cellcolor{c1} 2.67 & \cellcolor{c3} 0.22 & \cellcolor{c2} 0.6 & \cellcolor{c1} 1.75 & \cellcolor{c2} 0.54 & \cellcolor{c3} -0.26 & \cellcolor{c1} 3.0 & \cellcolor{c1} 1.87 & \cellcolor{c1} 1.78 & \cellcolor{c1} -2.17 & \cellcolor{c3} 0.3 & \cellcolor{c1} 1.9 & \cellcolor{c1} 1.45 & - & \cellcolor{c2} 0.42 & \cellcolor{c2} 0.97 & \cellcolor{c1} -1.49 \\ \ion{H}{$\alpha$} \scriptsize{(6563)} & \cellcolor{c2} -0.77 & - & \cellcolor{c3} -0.25 & \cellcolor{c1} 1.91 & \cellcolor{c2} -0.54 & \cellcolor{c3} -0.17 & \cellcolor{c2} 0.98 & \cellcolor{c3} -0.23 & \cellcolor{c1} -1.03 & \cellcolor{c1} 2.23 & \cellcolor{c1} 1.1 & \cellcolor{c1} 1.01 & \cellcolor{c1} -2.94 & \cellcolor{c2} -0.47 & \cellcolor{c1} 1.14 & \cellcolor{c2} 0.68 & \cellcolor{c2} -0.98 & \cellcolor{c2} -0.35 & \cellcolor{c3} 0.21 & \cellcolor{c1} -2.26 \\ \ion{S}{III} \scriptsize{(9069)} & \cellcolor{c2} -0.52 & \cellcolor{c3} 0.25 & - & \cellcolor{c1} 2.15 & - & \cellcolor{c3} 0.08 & \cellcolor{c1} 1.23 & - & \cellcolor{c2} -0.78 & \cellcolor{c1} 2.48 & \cellcolor{c1} 1.35 & \cellcolor{c1} 1.26 & \cellcolor{c1} -2.69 & \cellcolor{c3} -0.22 & \cellcolor{c1} 1.39 & \cellcolor{c2} 0.93 & \cellcolor{c2} -0.74 & - & \cellcolor{c2} 0.46 & \cellcolor{c1} -2.01 \\ \ion{S}{II} \scriptsize{(6731)} & \cellcolor{c1} -2.67 & \cellcolor{c1} -1.91 & \cellcolor{c1} -2.15 & - & \cellcolor{c1} -2.45 & \cellcolor{c1} -2.08 & \cellcolor{c2} -0.93 & \cellcolor{c1} -2.13 & \cellcolor{c1} -2.94 & \cellcolor{c2} 0.33 & \cellcolor{c2} -0.81 & \cellcolor{c2} -0.89 & \cellcolor{c1} -4.85 & \cellcolor{c1} -2.37 & - & \cellcolor{c1} -1.23 & \cellcolor{c1} -2.89 & \cellcolor{c1} -2.25 & \cellcolor{c1} -1.7 & \cellcolor{c1} -4.16 \\ \ion{He}{I} \scriptsize{(7065)} & \cellcolor{c3} -0.22 & \cellcolor{c2} 0.54 & - & \cellcolor{c1} 2.45 & - & \cellcolor{c2} 0.37 & \cellcolor{c1} 1.52 & - & \cellcolor{c2} -0.49 & \cellcolor{c1} 2.78 & \cellcolor{c1} 1.64 & \cellcolor{c1} 1.56 & \cellcolor{c1} -2.4 & - & \cellcolor{c1} 1.68 & \cellcolor{c1} 1.22 & \cellcolor{c2} -0.44 & - & \cellcolor{c2} 0.75 & \cellcolor{c1} -1.71 \\ \ion{Ar}{III} \scriptsize{(7136)} & \cellcolor{c2} -0.6 & \cellcolor{c3} 0.17 & \cellcolor{c3} -0.08 & \cellcolor{c1} 2.08 & \cellcolor{c2} -0.37 & - & \cellcolor{c1} 1.15 & \cellcolor{c3} -0.05 & \cellcolor{c2} -0.86 & \cellcolor{c1} 2.41 & \cellcolor{c1} 1.27 & \cellcolor{c1} 1.18 & \cellcolor{c1} -2.77 & \cellcolor{c3} -0.3 & \cellcolor{c1} 1.31 & \cellcolor{c2} 0.85 & \cellcolor{c2} -0.81 & \cellcolor{c3} -0.18 & \cellcolor{c2} 0.38 & \cellcolor{c1} -2.09 \\ \ion{N}{II} \scriptsize{(6583)} & \cellcolor{c1} -1.75 & \cellcolor{c2} -0.98 & \cellcolor{c1} -1.23 & \cellcolor{c2} 0.93 & \cellcolor{c1} -1.52 & \cellcolor{c1} -1.15 & - & \cellcolor{c1} -1.21 & \cellcolor{c1} -2.01 & \cellcolor{c1} 1.25 & \cellcolor{c3} 0.12 & \cellcolor{c3} 0.03 & \cellcolor{c1} -3.92 & \cellcolor{c1} -1.45 & \cellcolor{c3} 0.16 & - & \cellcolor{c1} -1.96 & \cellcolor{c1} -1.33 & \cellcolor{c2} -0.77 & \cellcolor{c1} -3.24 \\ \ion{O}{II} \scriptsize{(7320)} & \cellcolor{c2} -0.54 & \cellcolor{c3} 0.23 & - & \cellcolor{c1} 2.13 & - & \cellcolor{c3} 0.05 & \cellcolor{c1} 1.21 & - & \cellcolor{c2} -0.8 & \cellcolor{c1} 2.46 & \cellcolor{c1} 1.33 & \cellcolor{c1} 1.24 & \cellcolor{c1} -2.71 & \cellcolor{c3} -0.24 & \cellcolor{c1} 1.36 & \cellcolor{c2} 0.91 & \cellcolor{c2} -0.76 & - & \cellcolor{c2} 0.43 & \cellcolor{c1} -2.03 \\ \ion{Cl}{III} \scriptsize{(8434)} & \cellcolor{c3} 0.26 & \cellcolor{c1} 1.03 & \cellcolor{c2} 0.78 & \cellcolor{c1} 2.94 & \cellcolor{c2} 0.49 & \cellcolor{c2} 0.86 & \cellcolor{c1} 2.01 & \cellcolor{c2} 0.8 & - & \cellcolor{c1} 3.26 & \cellcolor{c1} 2.13 & \cellcolor{c1} 2.04 & \cellcolor{c1} -1.91 & \cellcolor{c2} 0.56 & \cellcolor{c1} 2.17 & \cellcolor{c1} 1.71 & \cellcolor{c3} 0.05 & \cellcolor{c2} 0.68 & \cellcolor{c1} 1.24 & \cellcolor{c1} -1.23 \\ \ion{Fe}{III} \scriptsize{(4986)} & \cellcolor{c1} -3.0 & \cellcolor{c1} -2.23 & \cellcolor{c1} -2.48 & \cellcolor{c2} -0.33 & \cellcolor{c1} -2.78 & \cellcolor{c1} -2.41 & \cellcolor{c1} -1.25 & \cellcolor{c1} -2.46 & \cellcolor{c1} -3.26 & - & \cellcolor{c1} -1.13 & \cellcolor{c1} -1.22 & \cellcolor{c1} -5.17 & \cellcolor{c1} -2.7 & \cellcolor{c1} -1.1 & \cellcolor{c1} -1.55 & \cellcolor{c1} -3.22 & \cellcolor{c1} -2.58 & \cellcolor{c1} -2.03 & \cellcolor{c1} -4.49 \\ \ion{Fe}{II} \scriptsize{(7155)} & \cellcolor{c1} -1.87 & \cellcolor{c1} -1.1 & \cellcolor{c1} -1.35 & \cellcolor{c2} 0.81 & \cellcolor{c1} -1.64 & \cellcolor{c1} -1.27 & \cellcolor{c3} -0.12 & \cellcolor{c1} -1.33 & \cellcolor{c1} -2.13 & \cellcolor{c1} 1.13 & - & \cellcolor{c3} -0.09 & \cellcolor{c1} -4.04 & \cellcolor{c1} -1.57 & \cellcolor{c3} 0.04 & \cellcolor{c2} -0.42 & \cellcolor{c1} -2.08 & \cellcolor{c1} -1.45 & \cellcolor{c2} -0.89 & \cellcolor{c1} -3.36 \\ \ion{Ni}{II} \scriptsize{(7378)} & \cellcolor{c1} -1.78 & \cellcolor{c1} -1.01 & \cellcolor{c1} -1.26 & \cellcolor{c2} 0.89 & \cellcolor{c1} -1.56 & \cellcolor{c1} -1.18 & \cellcolor{c3} -0.03 & \cellcolor{c1} -1.24 & \cellcolor{c1} -2.04 & \cellcolor{c1} 1.22 & \cellcolor{c3} 0.09 & - & \cellcolor{c1} -3.95 & \cellcolor{c1} -1.48 & \cellcolor{c3} 0.12 & \cellcolor{c2} -0.33 & \cellcolor{c1} -2.0 & \cellcolor{c1} -1.36 & \cellcolor{c2} -0.81 & \cellcolor{c1} -3.27 \\ \ion{Si}{III} \scriptsize{(8224)} & \cellcolor{c1} 2.17 & \cellcolor{c1} 2.94 & \cellcolor{c1} 2.69 & \cellcolor{c1} 4.85 & \cellcolor{c1} 2.4 & \cellcolor{c1} 2.77 & \cellcolor{c1} 3.92 & \cellcolor{c1} 2.71 & \cellcolor{c1} 1.91 & \cellcolor{c1} 5.17 & \cellcolor{c1} 4.04 & \cellcolor{c1} 3.95 & - & \cellcolor{c1} 2.47 & \cellcolor{c1} 4.08 & \cellcolor{c1} 3.62 & \cellcolor{c1} 1.96 & \cellcolor{c1} 2.59 & \cellcolor{c1} 3.15 & \cellcolor{c2} 0.68 \\ \ion{C}{II} \scriptsize{(5113)} & \cellcolor{c3} -0.3 & \cellcolor{c2} 0.47 & \cellcolor{c3} 0.22 & \cellcolor{c1} 2.37 & - & \cellcolor{c3} 0.3 & \cellcolor{c1} 1.45 & \cellcolor{c3} 0.24 & \cellcolor{c2} -0.56 & \cellcolor{c1} 2.7 & \cellcolor{c1} 1.57 & \cellcolor{c1} 1.48 & \cellcolor{c1} -2.47 & - & \cellcolor{c1} 1.6 & \cellcolor{c1} 1.15 & \cellcolor{c2} -0.52 & \cellcolor{c3} 0.12 & \cellcolor{c2} 0.67 & \cellcolor{c1} -1.79 \\ \ion{Cr}{II} \scriptsize{(8000)} & \cellcolor{c1} -1.9 & \cellcolor{c1} -1.14 & \cellcolor{c1} -1.39 & - & \cellcolor{c1} -1.68 & \cellcolor{c1} -1.31 & \cellcolor{c3} -0.16 & \cellcolor{c1} -1.36 & \cellcolor{c1} -2.17 & \cellcolor{c1} 1.1 & \cellcolor{c3} -0.04 & \cellcolor{c3} -0.12 & \cellcolor{c1} -4.08 & \cellcolor{c1} -1.6 & - & \cellcolor{c2} -0.46 & \cellcolor{c1} -2.12 & \cellcolor{c1} -1.49 & \cellcolor{c2} -0.93 & \cellcolor{c1} -3.39 \\ \ion{Cl}{II} \scriptsize{(8579)} & \cellcolor{c1} -1.45 & \cellcolor{c2} -0.68 & \cellcolor{c2} -0.93 & \cellcolor{c1} 1.23 & \cellcolor{c1} -1.22 & \cellcolor{c2} -0.85 & - & \cellcolor{c2} -0.91 & \cellcolor{c1} -1.71 & \cellcolor{c1} 1.55 & \cellcolor{c2} 0.42 & \cellcolor{c2} 0.33 & \cellcolor{c1} -3.62 & \cellcolor{c1} -1.15 & \cellcolor{c2} 0.46 & - & \cellcolor{c1} -1.66 & \cellcolor{c1} -1.03 & \cellcolor{c2} -0.47 & \cellcolor{c1} -2.94 \\ \ion{Al}{II} \scriptsize{(7064)} & - & \cellcolor{c2} 0.98 & \cellcolor{c2} 0.74 & \cellcolor{c1} 2.89 & \cellcolor{c2} 0.44 & \cellcolor{c2} 0.81 & \cellcolor{c1} 1.96 & \cellcolor{c2} 0.76 & \cellcolor{c3} -0.05 & \cellcolor{c1} 3.22 & \cellcolor{c1} 2.08 & \cellcolor{c1} 2.0 & \cellcolor{c1} -1.96 & \cellcolor{c2} 0.52 & \cellcolor{c1} 2.12 & \cellcolor{c1} 1.66 & - & \cellcolor{c2} 0.63 & \cellcolor{c1} 1.19 & \cellcolor{c1} -1.27 \\ \ion{Mg}{II} \scriptsize{(9218)} & \cellcolor{c2} -0.42 & \cellcolor{c2} 0.35 & - & \cellcolor{c1} 2.25 & - & \cellcolor{c3} 0.18 & \cellcolor{c1} 1.33 & - & \cellcolor{c2} -0.68 & \cellcolor{c1} 2.58 & \cellcolor{c1} 1.45 & \cellcolor{c1} 1.36 & \cellcolor{c1} -2.59 & \cellcolor{c3} -0.12 & \cellcolor{c1} 1.49 & \cellcolor{c1} 1.03 & \cellcolor{c2} -0.63 & - & \cellcolor{c2} 0.56 & \cellcolor{c1} -1.91 \\ \ion{P}{II} \scriptsize{(7876)} & \cellcolor{c2} -0.97 & \cellcolor{c3} -0.21 & \cellcolor{c2} -0.46 & \cellcolor{c1} 1.7 & \cellcolor{c2} -0.75 & \cellcolor{c2} -0.38 & \cellcolor{c2} 0.77 & \cellcolor{c2} -0.43 & \cellcolor{c1} -1.24 & \cellcolor{c1} 2.03 & \cellcolor{c2} 0.89 & \cellcolor{c2} 0.81 & \cellcolor{c1} -3.15 & \cellcolor{c2} -0.67 & \cellcolor{c2} 0.93 & \cellcolor{c2} 0.47 & \cellcolor{c1} -1.19 & \cellcolor{c2} -0.56 & - & \cellcolor{c1} -2.46 \\ \ion{B}{II} \scriptsize{(7030)} & \cellcolor{c1} 1.49 & \cellcolor{c1} 2.26 & \cellcolor{c1} 2.01 & \cellcolor{c1} 4.16 & \cellcolor{c1} 1.71 & \cellcolor{c1} 2.09 & \cellcolor{c1} 3.24 & \cellcolor{c1} 2.03 & \cellcolor{c1} 1.23 & \cellcolor{c1} 4.49 & \cellcolor{c1} 3.36 & \cellcolor{c1} 3.27 & \cellcolor{c2} -0.68 & \cellcolor{c1} 1.79 & \cellcolor{c1} 3.39 & \cellcolor{c1} 2.94 & \cellcolor{c1} 1.27 & \cellcolor{c1} 1.91 & \cellcolor{c1} 2.46 & - \\  \hline $\bar{l}/\bar{l}_{\text{H}\alpha}$ & 1.4 & 1.0 & 0.14 & 0.01 & 0.02 & 0.08 & 0.17 & 0.04 & $10^{-4}$ & $10^{-5}$ & $10^{-3}$ & $10^{-3}$ & $10^{-20}$ & $10^{-5}$ & $10^{-5}$ & $10^{-3}$ & $10^{-9}$ & $10^{-6}$ & $10^{-4}$ & $10^{-20}$ \\ 
        \hline
    \end{tabular}
    \caption{ A line ratio sensitivity Table, including lines selected by the line-selection procedure over the MUSE wavelength range 4800 \AA\ to 9300 \AA. The lines are ordered from left to right (and top to bottom) by decreasing magnitude of the line measure. Therefore, line ratios towards the upper left hand corner are on average both brighter and more sensitive to $G$. The bottom row shows the mean luminosity of the line $\bar{l}$ over the mean H$\alpha$ luminosity $\bar{l}_{\text{H}\alpha}$. The values in the Table are $\log_{10}(f_{\text{ij}})$ if $|R_{\text{ij}}| \ge 0.7$ and dashed if $|R_{\text{ij}}| < 0.7$. The value $f_{\text{ij}}$ is the factor by which the line ratio between lines i and j changes from $d=0.05$ pc to $d=1.41$ pc (or $10^3$ G$_\text{0}$ \text{--} $10^6$ G$_\text{0}$), and $R_{\text{ij}}$ is the trends correlation. The colour coding is described by equation \ref{colours}.}
    \label{tab:line_ratios_MUSE}
\end{table*}
}

{
\setlength{\tabcolsep}{1.9pt}
\renewcommand{\arraystretch}{1.5}
\begin{table*}
    \centering
    \begin{tabular}{c|ccccccccccccccccccccc}
        \hline 
        species & \ion{O}{III} & \ion{He}{I} & \ion{H}{$\alpha$} & \ion{C}{III} & \ion{Si}{III} & \ion{Ne}{III} & \ion{S}{III} & \ion{S}{II} & \ion{O}{II} & \ion{N}{II} & \ion{N}{III} & \ion{Fe}{II} & \ion{Al}{II} & \ion{Cl}{III} & \ion{C}{II} & \ion{Ar}{III} & \ion{Fe}{III} & \ion{Ne}{II} & \ion{Ni}{II} & \ion{Ar}{II} \\ ($\lambda$ [\AA]) & \scriptsize{(5007)} & \scriptsize{(1203)} & \scriptsize{(6563)} & \scriptsize{(1909)} & \scriptsize{(1295)} & \scriptsize{(3869)} & \scriptsize{(9531)} & \scriptsize{(6731)} & \scriptsize{(3729)} & \scriptsize{(6583)} & \scriptsize{(1750)} & \scriptsize{(5.339)} & \scriptsize{(1760)} & \scriptsize{(3343)} & \scriptsize{(1335)} & \scriptsize{(7136)} & \scriptsize{(2438)} & \scriptsize{(12.81)} & \scriptsize{(1.191)} & \scriptsize{(6.983)} \\  \hline \ion{O}{III} \scriptsize{(5007)} & - & \cellcolor{c2} 0.56 & \cellcolor{c2} 0.77 & - & \cellcolor{c1} -2.15 & - & \cellcolor{c2} 0.52 & \cellcolor{c1} 2.67 & \cellcolor{c1} 2.25 & \cellcolor{c1} 1.75 & \cellcolor{c3} -0.26 & \cellcolor{c1} 3.68 & \cellcolor{c2} -0.85 & \cellcolor{c3} -0.26 & \cellcolor{c2} 0.44 & \cellcolor{c2} 0.6 & \cellcolor{c1} -1.11 & \cellcolor{c1} 1.2 & \cellcolor{c1} 3.54 & \cellcolor{c1} 1.59 \\ \ion{He}{I} \scriptsize{(1203)} & \cellcolor{c2} -0.56 & - & \cellcolor{c3} 0.21 & \cellcolor{c2} -0.71 & \cellcolor{c1} -2.71 & \cellcolor{c2} -0.65 & \cellcolor{c3} -0.04 & \cellcolor{c1} 2.12 & \cellcolor{c1} 1.7 & \cellcolor{c1} 1.19 & \cellcolor{c2} -0.81 & \cellcolor{c1} 3.13 & \cellcolor{c1} -1.41 & \cellcolor{c2} -0.82 & \cellcolor{c3} -0.11 & \cellcolor{c3} 0.04 & \cellcolor{c1} -1.67 & \cellcolor{c2} 0.65 & \cellcolor{c1} 2.99 & \cellcolor{c1} 1.04 \\ \ion{H}{$\alpha$} \scriptsize{(6563)} & \cellcolor{c2} -0.77 & \cellcolor{c3} -0.21 & - & \cellcolor{c2} -0.92 & \cellcolor{c1} -2.92 & \cellcolor{c2} -0.86 & \cellcolor{c3} -0.25 & \cellcolor{c1} 1.91 & \cellcolor{c1} 1.49 & \cellcolor{c2} 0.98 & \cellcolor{c1} -1.03 & \cellcolor{c1} 2.91 & \cellcolor{c1} -1.62 & \cellcolor{c1} -1.03 & \cellcolor{c2} -0.33 & \cellcolor{c3} -0.17 & \cellcolor{c1} -1.88 & \cellcolor{c2} 0.43 & \cellcolor{c1} 2.78 & \cellcolor{c2} 0.82 \\ \ion{C}{III} \scriptsize{(1909)} & - & \cellcolor{c2} 0.71 & \cellcolor{c2} 0.92 & - & \cellcolor{c1} -2.0 & - & \cellcolor{c2} 0.68 & \cellcolor{c1} 2.83 & \cellcolor{c1} 2.41 & \cellcolor{c1} 1.9 & \cellcolor{c3} -0.1 & \cellcolor{c1} 3.84 & \cellcolor{c2} -0.7 & \cellcolor{c3} -0.11 & \cellcolor{c2} 0.6 & \cellcolor{c2} 0.75 & \cellcolor{c2} -0.95 & \cellcolor{c1} 1.36 & \cellcolor{c1} 3.7 & \cellcolor{c1} 1.75 \\ \ion{Si}{III} \scriptsize{(1295)} & \cellcolor{c1} 2.15 & \cellcolor{c1} 2.71 & \cellcolor{c1} 2.92 & \cellcolor{c1} 2.0 & - & \cellcolor{c1} 2.06 & \cellcolor{c1} 2.67 & \cellcolor{c1} 4.83 & \cellcolor{c1} 4.41 & \cellcolor{c1} 3.9 & \cellcolor{c1} 1.9 & \cellcolor{c1} 5.83 & \cellcolor{c1} 1.3 & \cellcolor{c1} 1.89 & \cellcolor{c1} 2.59 & \cellcolor{c1} 2.75 & \cellcolor{c1} 1.04 & \cellcolor{c1} 3.35 & \cellcolor{c1} 5.7 & \cellcolor{c1} 3.74 \\ \ion{Ne}{III} \scriptsize{(3869)} & - & \cellcolor{c2} 0.65 & \cellcolor{c2} 0.86 & - & \cellcolor{c1} -2.06 & - & \cellcolor{c2} 0.61 & \cellcolor{c1} 2.77 & \cellcolor{c1} 2.35 & \cellcolor{c1} 1.84 & \cellcolor{c3} -0.16 & \cellcolor{c1} 3.78 & \cellcolor{c2} -0.76 & \cellcolor{c3} -0.17 & \cellcolor{c2} 0.54 & \cellcolor{c2} 0.69 & \cellcolor{c1} -1.02 & \cellcolor{c1} 1.3 & \cellcolor{c1} 3.64 & \cellcolor{c1} 1.69 \\ \ion{S}{III} \scriptsize{(9531)} & \cellcolor{c2} -0.52 & \cellcolor{c3} 0.04 & \cellcolor{c3} 0.25 & \cellcolor{c2} -0.68 & \cellcolor{c1} -2.67 & \cellcolor{c2} -0.61 & - & \cellcolor{c1} 2.15 & \cellcolor{c1} 1.73 & \cellcolor{c1} 1.23 & \cellcolor{c2} -0.78 & \cellcolor{c1} 3.16 & \cellcolor{c1} -1.37 & \cellcolor{c2} -0.78 & - & \cellcolor{c3} 0.08 & \cellcolor{c1} -1.63 & \cellcolor{c2} 0.68 & \cellcolor{c1} 3.02 & \cellcolor{c1} 1.07 \\ \ion{S}{II} \scriptsize{(6731)} & \cellcolor{c1} -2.67 & \cellcolor{c1} -2.12 & \cellcolor{c1} -1.91 & \cellcolor{c1} -2.83 & \cellcolor{c1} -4.83 & \cellcolor{c1} -2.77 & \cellcolor{c1} -2.15 & - & \cellcolor{c2} -0.42 & \cellcolor{c2} -0.93 & \cellcolor{c1} -2.93 & \cellcolor{c1} 1.01 & \cellcolor{c1} -3.53 & \cellcolor{c1} -2.94 & \cellcolor{c1} -2.23 & \cellcolor{c1} -2.08 & \cellcolor{c1} -3.78 & \cellcolor{c1} -1.47 & \cellcolor{c2} 0.87 & \cellcolor{c1} -1.08 \\ \ion{O}{II} \scriptsize{(3729)} & \cellcolor{c1} -2.25 & \cellcolor{c1} -1.7 & \cellcolor{c1} -1.49 & \cellcolor{c1} -2.41 & \cellcolor{c1} -4.41 & \cellcolor{c1} -2.35 & \cellcolor{c1} -1.73 & \cellcolor{c2} 0.42 & - & \cellcolor{c2} -0.51 & \cellcolor{c1} -2.51 & \cellcolor{c1} 1.43 & \cellcolor{c1} -3.11 & \cellcolor{c1} -2.52 & \cellcolor{c1} -1.81 & \cellcolor{c1} -1.66 & \cellcolor{c1} -3.36 & \cellcolor{c1} -1.05 & \cellcolor{c1} 1.29 & \cellcolor{c2} -0.66 \\ \ion{N}{II} \scriptsize{(6583)} & \cellcolor{c1} -1.75 & \cellcolor{c1} -1.19 & \cellcolor{c2} -0.98 & \cellcolor{c1} -1.9 & \cellcolor{c1} -3.9 & \cellcolor{c1} -1.84 & \cellcolor{c1} -1.23 & \cellcolor{c2} 0.93 & \cellcolor{c2} 0.51 & - & \cellcolor{c1} -2.0 & \cellcolor{c1} 1.94 & \cellcolor{c1} -2.6 & \cellcolor{c1} -2.01 & \cellcolor{c1} -1.3 & \cellcolor{c1} -1.15 & \cellcolor{c1} -2.86 & \cellcolor{c2} -0.54 & \cellcolor{c1} 1.8 & - \\ \ion{N}{III} \scriptsize{(1750)} & \cellcolor{c3} 0.26 & \cellcolor{c2} 0.81 & \cellcolor{c1} 1.03 & \cellcolor{c3} 0.1 & \cellcolor{c1} -1.9 & \cellcolor{c3} 0.16 & \cellcolor{c2} 0.78 & \cellcolor{c1} 2.93 & \cellcolor{c1} 2.51 & \cellcolor{c1} 2.0 & - & \cellcolor{c1} 3.94 & \cellcolor{c2} -0.59 & - & \cellcolor{c2} 0.7 & \cellcolor{c2} 0.85 & \cellcolor{c2} -0.85 & \cellcolor{c1} 1.46 & \cellcolor{c1} 3.8 & \cellcolor{c1} 1.85 \\ \ion{Fe}{II} \scriptsize{(5.339)} & \cellcolor{c1} -3.68 & \cellcolor{c1} -3.13 & \cellcolor{c1} -2.91 & \cellcolor{c1} -3.84 & \cellcolor{c1} -5.83 & \cellcolor{c1} -3.78 & \cellcolor{c1} -3.16 & \cellcolor{c1} -1.01 & \cellcolor{c1} -1.43 & \cellcolor{c1} -1.94 & \cellcolor{c1} -3.94 & - & \cellcolor{c1} -4.53 & \cellcolor{c1} -3.95 & \cellcolor{c1} -3.24 & \cellcolor{c1} -3.09 & \cellcolor{c1} -4.79 & \cellcolor{c1} -2.48 & \cellcolor{c3} -0.14 & \cellcolor{c1} -2.09 \\ \ion{Al}{II} \scriptsize{(1760)} & \cellcolor{c2} 0.85 & \cellcolor{c1} 1.41 & \cellcolor{c1} 1.62 & \cellcolor{c2} 0.7 & \cellcolor{c1} -1.3 & \cellcolor{c2} 0.76 & \cellcolor{c1} 1.37 & \cellcolor{c1} 3.53 & \cellcolor{c1} 3.11 & \cellcolor{c1} 2.6 & \cellcolor{c2} 0.59 & \cellcolor{c1} 4.53 & - & \cellcolor{c2} 0.59 & \cellcolor{c1} 1.29 & \cellcolor{c1} 1.45 & \cellcolor{c3} -0.26 & \cellcolor{c1} 2.05 & \cellcolor{c1} 4.4 & \cellcolor{c1} 2.44 \\ \ion{Cl}{III} \scriptsize{(3343)} & \cellcolor{c3} 0.26 & \cellcolor{c2} 0.82 & \cellcolor{c1} 1.03 & \cellcolor{c3} 0.11 & \cellcolor{c1} -1.89 & \cellcolor{c3} 0.17 & \cellcolor{c2} 0.78 & \cellcolor{c1} 2.94 & \cellcolor{c1} 2.52 & \cellcolor{c1} 2.01 & - & \cellcolor{c1} 3.95 & \cellcolor{c2} -0.59 & - & \cellcolor{c2} 0.71 & \cellcolor{c2} 0.86 & \cellcolor{c2} -0.85 & \cellcolor{c1} 1.47 & \cellcolor{c1} 3.81 & \cellcolor{c1} 1.86 \\ \ion{C}{II} \scriptsize{(1335)} & \cellcolor{c2} -0.44 & \cellcolor{c3} 0.11 & \cellcolor{c2} 0.33 & \cellcolor{c2} -0.6 & \cellcolor{c1} -2.59 & \cellcolor{c2} -0.54 & - & \cellcolor{c1} 2.23 & \cellcolor{c1} 1.81 & \cellcolor{c1} 1.3 & \cellcolor{c2} -0.7 & \cellcolor{c1} 3.24 & \cellcolor{c1} -1.29 & \cellcolor{c2} -0.71 & - & \cellcolor{c3} 0.15 & \cellcolor{c1} -1.55 & \cellcolor{c2} 0.76 & \cellcolor{c1} 3.1 & \cellcolor{c1} 1.15 \\ \ion{Ar}{III} \scriptsize{(7136)} & \cellcolor{c2} -0.6 & \cellcolor{c3} -0.04 & \cellcolor{c3} 0.17 & \cellcolor{c2} -0.75 & \cellcolor{c1} -2.75 & \cellcolor{c2} -0.69 & \cellcolor{c3} -0.08 & \cellcolor{c1} 2.08 & \cellcolor{c1} 1.66 & \cellcolor{c1} 1.15 & \cellcolor{c2} -0.85 & \cellcolor{c1} 3.09 & \cellcolor{c1} -1.45 & \cellcolor{c2} -0.86 & \cellcolor{c3} -0.15 & - & \cellcolor{c1} -1.71 & \cellcolor{c2} 0.61 & \cellcolor{c1} 2.95 & \cellcolor{c2} 1.0 \\ \ion{Fe}{III} \scriptsize{(2438)} & \cellcolor{c1} 1.11 & \cellcolor{c1} 1.67 & \cellcolor{c1} 1.88 & \cellcolor{c2} 0.95 & \cellcolor{c1} -1.04 & \cellcolor{c1} 1.02 & \cellcolor{c1} 1.63 & \cellcolor{c1} 3.78 & \cellcolor{c1} 3.36 & \cellcolor{c1} 2.86 & \cellcolor{c2} 0.85 & \cellcolor{c1} 4.79 & \cellcolor{c3} 0.26 & \cellcolor{c2} 0.85 & \cellcolor{c1} 1.55 & \cellcolor{c1} 1.71 & - & \cellcolor{c1} 2.31 & \cellcolor{c1} 4.66 & \cellcolor{c1} 2.7 \\ \ion{Ne}{II} \scriptsize{(12.81)} & \cellcolor{c1} -1.2 & \cellcolor{c2} -0.65 & \cellcolor{c2} -0.43 & \cellcolor{c1} -1.36 & \cellcolor{c1} -3.35 & \cellcolor{c1} -1.3 & \cellcolor{c2} -0.68 & \cellcolor{c1} 1.47 & \cellcolor{c1} 1.05 & \cellcolor{c2} 0.54 & \cellcolor{c1} -1.46 & \cellcolor{c1} 2.48 & \cellcolor{c1} -2.05 & \cellcolor{c1} -1.47 & \cellcolor{c2} -0.76 & \cellcolor{c2} -0.61 & \cellcolor{c1} -2.31 & - & \cellcolor{c1} 2.34 & \cellcolor{c2} 0.39 \\ \ion{Ni}{II} \scriptsize{(1.191)} & \cellcolor{c1} -3.54 & \cellcolor{c1} -2.99 & \cellcolor{c1} -2.78 & \cellcolor{c1} -3.7 & \cellcolor{c1} -5.7 & \cellcolor{c1} -3.64 & \cellcolor{c1} -3.02 & \cellcolor{c2} -0.87 & \cellcolor{c1} -1.29 & \cellcolor{c1} -1.8 & \cellcolor{c1} -3.8 & \cellcolor{c3} 0.14 & \cellcolor{c1} -4.4 & \cellcolor{c1} -3.81 & \cellcolor{c1} -3.1 & \cellcolor{c1} -2.95 & \cellcolor{c1} -4.66 & \cellcolor{c1} -2.34 & - & \cellcolor{c1} -1.95 \\ \ion{Ar}{II} \scriptsize{(6.983)} & \cellcolor{c1} -1.59 & \cellcolor{c1} -1.04 & \cellcolor{c2} -0.82 & \cellcolor{c1} -1.75 & \cellcolor{c1} -3.74 & \cellcolor{c1} -1.69 & \cellcolor{c1} -1.07 & \cellcolor{c1} 1.08 & \cellcolor{c2} 0.66 & - & \cellcolor{c1} -1.85 & \cellcolor{c1} 2.09 & \cellcolor{c1} -2.44 & \cellcolor{c1} -1.86 & \cellcolor{c1} -1.15 & \cellcolor{c2} -1.0 & \cellcolor{c1} -2.7 & \cellcolor{c2} -0.39 & \cellcolor{c1} 1.95 & - \\  \hline $\bar{l}/\bar{l}_{\text{H}\alpha}$ & 1.4 & 1.5 & 1.0 & 0.06 & $10^{-5}$ & 0.04 & 0.35 & 0.01 & 0.05 & 0.17 & $10^{-3}$ & $10^{-4}$ & $10^{-6}$ & $10^{-3}$ & 0.05 & 0.08 & $10^{-7}$ & 0.07 & $10^{-6}$ & $10^{-2}$ \\ 
        \hline
    \end{tabular}
    \caption{A line ratio sensitivity table, including lines selected over the wavelength range 1,150 \AA\ to 28.5 $\mu$m. This corresponds to the lower bound of the UVEX wavelength range and the upper bound of JWST. Wavelengths longer than 10,000 \AA\ are given in $\mu$m, whilst those below are in \AA. The table is constructed identically to Table \ref{tab:line_ratios_MUSE}, a summary of which is given in its caption.}
    \label{tab:line_ratios_ALL}
\end{table*}
}

\subsubsection{Comparison to observations}
 \label{sec:obsCompare}
 The line ratio [\ion{N}{II}] 6583 \AA\ / [\ion{S}{II}] 6731+6716 \AA\ has been proposed by \cite{2025A&A...693A..87M} as a diagnostic of exposure to high FUV fields. This is consistent with the trend that our models predict for [\ion{N}{II}] 6583 \AA\ / [\ion{S}{II}] 6731 \AA\, (note that we neglect the contribution from [\ion{S}{II}] 6716 \AA\ since its dimmer and behaves similarly to [\ion{S}{II}] 6731 \AA\ anyway).

 \cite{2025A&A...693A..87M} also proposed that the ratio [\ion{O}{I}] 6300 \AA\ / [\ion{S}{II}] 6731+6716 \AA\ follows the inverse relation, decreasing with increasing FUV field strength. In our line ratio analysis we neglected [\ion{O}{I}] 6300 \AA\ emission since it originates from regions other than the ionized outflow \citep{1998ApJ...502L..71S}. However, theoretical models by \cite{2023MNRAS.518.5563B} show that [\ion{O}{I}] 6300 \AA\ emission increases with $G$, whilst our model shows that the relative contribution from [\ion{S}{II}] 6731 \AA\ decreases with increasing $G$. In all cases but one in Tables \ref{tab:line_ratios_MUSE} and \ref{tab:line_ratios_ALL} ratios with [\ion{S}{II}] 6731 \AA\ on the numerator fall at high FUV fields and rise when [\ion{S}{II}] 6731 \AA\ is on the denominator. This would suggest that the ratio [\ion{O}{I}] 6300 \AA\ / [\ion{S}{II}] 6731+6716 \AA\ actually increases with increasing $G$. However, a model including all sources of [\ion{O}{I}] 6300 \AA\ emission, and which models the ionization front in more detail would be required to verify this. Further observations of the behaviour of this ratio in clusters will also strengthen the empirical evidence for a relation. 
 
We undertake a preliminary test of the [\ion{S}{II}] 6731 \AA\ / [\ion{O}{III}] 4959 \AA\ line ratio trend in MUSE observations of the ONC proplyds. Emission from \ion{O}{III} always obeys $j_{5007} \approx 3j_{4959}$, so both follow the same trends with the same $f$. For each line we consider the intensity observed along the line of sight from the proplyd to the UV source and integrate it over a semicircle (the face of the proplyd) to calculate the luminosity. The line ratio is shown for five proplyds in Table \ref{tab:SII/OIII_observations}. Images of all five can be seen in Figure 1 of \cite{2024A&A...687A..93A} in multiple emission lines. Although this is only a small sample size, a preliminary trend is present, albeit with an anomalous result from 170-337. The excess [\ion{S}{II}] 6731 \AA\ emission in 170-337 probably has a large contribution from a jet \citep{2022MNRAS.514..744M}. Additionally, 170-337 may be either far in the background of foreground, i.e. $d >> d_\perp$. The lack of an obscuring disc/PDR in Figure 1 of \cite{2024A&A...687A..93A} suggests that it is in the background. The projected length of the tail is $L_{\text{tail},\perp} \approx 2r_{\text{IF}}$, while the true length by theory is $L_{\text{tail}}\approx 5r_{\text{IF}}$ \citep{1998ApJ...499..758J}. This suggests that the true distance from the proplyd is $d\approx 2.5d_{\perp}=0.078$ pc. 170-337 being at this distance does not quite make its line ratio fit the trend, but does provide a partial explanation, alongside the jet, for the disparity. The remaining proplyds in Figure 1 of \cite{2024A&A...687A..93A}, 170-249, 173-236, and 177-341W all have tails much closer to $5r_{IF}$. This suggests they are nearer the plane of the image and the physical distance is closer to the projected distance.

{
\renewcommand{\arraystretch}{1.5}
\begin{table}
    \centering
    \begin{tabular}{ccc}
        \hline
        Proplyd & Projected Distance $d_\perp$ [pc] & [\ion{S}{II}] 6731 \AA\ / [\ion{O}{III}] 4959 \AA\ \\
        170-337 & 0.031 & 630 \\
        177-341W & 0.049 & 0.0095 \\
        159-350 & 0.056 & 0.014 \\
        170-249 & 0.068 & 0.17 \\
        173-236 & 0.095 & 19 \\
        \hline
    \end{tabular}
    \caption{The line ratio [\ion{S}{II}] 6731 \AA\ / [\ion{O}{III}] 4959 \AA\ calculated for a subset of the proplyds and their projected distances \citep{2024A&A...687A..93A}. The line ratio increases with projected distance, as expected, bar proplyd 170-337 which likely has a jet.}
\label{tab:SII/OIII_observations}
\end{table}
}

\subsection{Physical explanation for line ratio variations}\label{Physical Explanation}

CELs are produced when a free electron collisionally excites a bound electron to a higher energy state, followed by a radiative decay. The emissivity of the CEL is $j=n_\text{2}A_{\text{21}}E_{21}$, where $n_\text{2}$ is the number density of the excited state, $A_{\text{21}}$ is the radiative decay rate, and $E_{21}$ is the energy difference between the excited and ground state. The population of the excited state is a function of both densities and temperature. The total luminosity is found by integrating the emissivity over the volume of the emitting region. We discuss which of these physical variables most affect the proplyd line ratios in the following subsections.

\subsubsection{Volume of emission}

Every metal has successive ionization fronts, from their neutral state to highly ionized in the outer wind. In order to understand how the radius of a metal's ionization front scales with the radius of the hydrogen ionization front, we consider the global ionization parameter
\begin{equation}
    U_\text{IF}=\frac{\phi(H)}{n_\text{IF}c},
\end{equation}
where $\phi(H)$ is the intensity of ionizing photons incident on the proplyd and we use the density at the hydrogen ionization front $n_\text{IF}$ to characterise the density scale of the system. 

\begin{figure}
    \centering
    \includegraphics[width=\columnwidth]{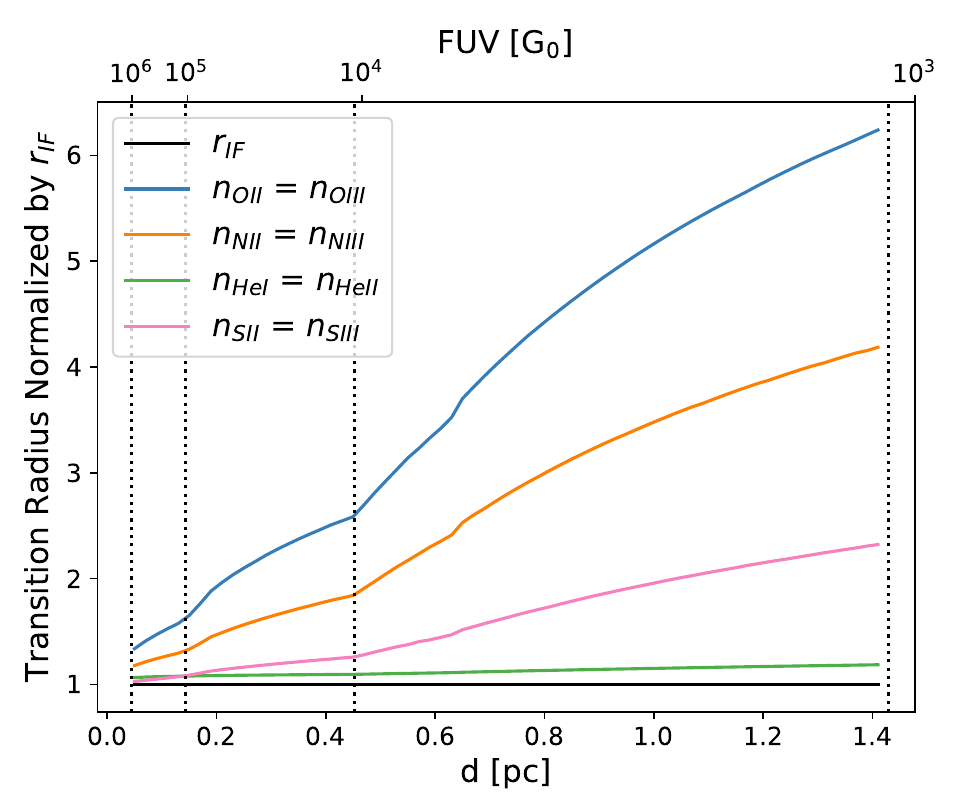} 
    \caption{Transition radii (I-fronts) as a function of distance between the base-case proplyd and the UV source, $d=0.05$ pc -- $d=1.41$ pc, for the transitions: \ion{O}{II}-\ion{O}{III}, \ion{N}{II}-\ion{N}{III}, \ion{He}{I}-\ion{He}{II}, \ion{S}{II}-\ion{S}{III}. These are normalized by the radius of the hydrogen ionization front, hence the $r_{\text{IF}}$ line is always 1.}
    \label{fig:I_fronts}
\end{figure}

In our model, any proplyd irradiated by a given UV source is entirely defined by $d$ and $\dot{M}$. We consider the transformation of the systems properties by $d_1= Yd_0$ and $\dot{M}_1 = X\dot{M}_0$. The untransformed and transformed properties are denoted by $0$ and $1$, respectively. The density at the hydrogen ionization front is $n_{\text{IF}}=C\dot{M}/r_{\text{IF}}^2$, where $C$ has absorbed the constants. The ratio of the hydrogen ionization front density between the transformed and untransformed case is 
\begin{equation}
    \frac{n_{\text{IF},1}}{n_{\text{IF},0}} = \frac{\dot{M}_1}{\dot{M}_0} \frac{r_{\text{IF},0}^2}{r_{\text{IF},1}^2} = \frac{1}{X^{1/3}Y^{4/3}},   
\end{equation}
where we have used that $r_{\text{IF},1}=r_{\text{IF},0}X^{2/3}Y^{2/3}$ (which can be obtained from equation \ref{Ifront}). The intensity scales simply as $\Phi_1=\Phi_0Y^{-2}$. Putting this all together, we get the scaling that $U_\text{IF,1} = U_\text{IF,0}X^{1/3}Y^{-2/3}$. We assume that out in the ionized wind where metal ionization fronts occur that $\phi(H)$ is approximately constant and therefore that $U_\text{IF}n_\text{IF}=U_\text{m}n_\text{m}$. The ionization front of a metal m occurs at its characteristic ionization parameter $U_\text{m}$ where the total density is $n_\text{m}$, i.e. $U_\text{m,1}=U_\text{m,0}$. We can then form the ratio
\begin{equation}
    \frac{U_\text{IF,1}}{U_\text{IF,0}} = \frac{n_\text{m,1}}{n_\text{IF,1}} \frac{n_\text{IF,0}}{n_\text{m,0}}= \left( \frac{r_\text{IF,1}}{r_\text{m,1}} \right)^2 \left( \frac{r_\text{m,0}}{r_\text{IF,0}} \right)^2,
\end{equation}
which can be rearranged to show how the metal ionization front scales relative to the hydrogen ionization front as
\begin{equation}
    \frac{r_\text{m,1}}{r_\text{IF,1}} = \frac{r_\text{m,0}}{r_\text{IF,0}}\frac{X^{1/6}}{Y^{1/3}}.
\end{equation}
If we assume that proplyd 1 is farther from the UV source than proplyd 0 then $Y>1$ and $X<1$, hence it is always the case that $r_\text{m,1} / r_\text{IF,1} > r_\text{m,0} / r_\text{IF,0}$. For the base-case proplyd we plot in Figure \ref{fig:I_fronts} $r_\text{m} / r_\text{IF}$ as a function of distance from the UV source for the ionization fronts: \ion{He}{I}/\ion{He}{II}, \ion{S}{II}/\ion{S}{III}, \ion{O}{II}/\ion{O}{III}, and \ion{N}{II}/\ion{N}{III}.

From the trends in $r_\text{m} / r_\text{IF}$ it is apparent why the line ratio [\ion{S}{II}] 6731 \AA\ / [\ion{O}{III}] 4959 \AA\ is such an effective diagnostic. Between $d= 0.05$ pc and $1.41$ pc the radii over which \ion{S}{II} and \ion{O}{III} are the dominant ionization state falls by a factor of $\approx25$ and $\approx7$, respectively. The relative contribution from [\ion{O}{III}] 4959 \AA\ emission increases whilst that from [\ion{S}{II}] 6731 \AA\ decreases. Both act to increase the ratio [\ion{S}{II}] 6731 \AA\ / [\ion{O}{III}] 4959 \AA\ for small $d$. For this reason strong diagnostics often come from a line ratio where one species is in ionization state II whilst the other is in III. We see in Figure \ref{fig:Best_Line_Ratios} that of the five ratios shown, the four most sensitive satisfy this, whilst the least sensitive does not.

\subsubsection{Density in the outflow}\label{Density in the Outflow}

The critical density of the excited state affects the sensitivity of the line ratio. Excited states which have critical densities $n_\text{crit}<n_\textrm{IF}$ have emissivities that scale as $j \propto n$, whilst when $n_\text{crit}>n_\textrm{IF}$ then $j \propto n^2$. Therefore, line ratios $l_\text{i}/l_\text{j}$ where one line has $n_\text{crit,i}<n_\textrm{IF}$ and one has $n_\text{crit,j}>n_\textrm{IF}$ depend on the density in the flow. 

For example, in Tables \ref{tab:line_ratios_MUSE} and \ref{tab:line_ratios_ALL} we see ratios of the lines [\ion{O}{II}] 3726 \AA\ and [\ion{O}{II}] 7320 \AA\ with the line [\ion{O}{III}] 5007 \AA. For the upper state of each line the critical electron density $n_{\textrm{crit}}$ at $10^4$ K and the energy $E_{\textrm{u}}$ of the state are shown in table \ref{tab:line_properties}. For the base-case proplyd, between $10^3$ and $10^6$ G$_0$ the electron density at the hydrogen ionization front is between $\approx 2\times10^4$ and $2\times10^5$ cm$^{-3}$. Therefore, the emissivities of the lines follow the scaling: $j_\text{3726}\propto n$, $j_\text{7320}\propto n^2$, and $j_\text{4959}\propto n^2$. Since the lines in the ratio [\ion{O}{II}] 7320 \AA\ / [\ion{O}{III}] 5007 \AA\ have the same density scaling the density in the flow is inconsequential. However, for the line ratio [\ion{O}{II}] 3726 \AA\ / [\ion{O}{III}] 5007 \AA\ the density matters. Firstly, for increasing $d$, the relative volume of \ion{O}{II} emission rises, and the relative volume of \ion{O}{III} emission falls. Secondly, the density in the ionized wind decreases and the emissivity $j_\text{4959}$ falls more than $j_\text{3726}$ does. Both effects result in the line ratio [\ion{O}{II}] 3726 \AA\ / [\ion{O}{III}] 5007 \AA\ rising with the distance from the UV source.

{
\setlength{\tabcolsep}{12pt}
\renewcommand{\arraystretch}{1.5}
\begin{table}
    \centering
    \begin{tabular}{lcr}
    \hline
        Line [\AA] & $n_{\textrm{crit}}$ [cm$^{-3}$] & $E_{\textrm{u}}$ [eV] \\
        \ion{O}{II} 3726 & $4.5 \times10^3$ & 3.3 \\
        \ion{O}{II} 7320 & $1.0 \times10^7$ & 5.0 \\
        \ion{O}{III} 5007 & $6.4 \times10^5$ & 2.5 \\
        \ion{S}{II} 6731 & $1.5 \times10^4$ & 1.8 \\ 
    \hline
    \end{tabular}
    \caption{For emission lines of interest we show the critical density $n_{\textrm{crit}}$ \citep{2011piim.book.....D} and the energy of the excited state $E_{\textrm{u}}$ \citep{NIST_ASD}.}
    \label{tab:line_properties}
\end{table}
}

\subsubsection{Temperature in the outflow}\label{Temperature in the Outflow}

A generally lower order but sometimes still significant effect is the dependence of the temperature on the population of the excited states. For a CEL this depends on the energy of the excited state $E$ since the emissivity $j \propto e^{-E/{k_{\textrm{B}}T}}$. For the base-case proplyd, between $10^3$ and $10^6$ G$_\text{0}$ the peak temperature rises from $\approx9\times 10^3-$  $1.1\times10^4$ K. With only knowledge of the wavelength, we can say that $E$ is at least as large as the photon energy. However, longer wavelength lines can still come from high-energy states (decaying to intermediate states), so the energy levels should be considered.

For a 3 level system the nebular line is from state 2 to 1 while the auroral line is from state 3 to 2. Due to its higher energy the population of the excited auroral line state is more sensitive to temperature than the excited nebular state. The higher energy also means the auroral lines excited state is less populated and the line will be weaker. Therefore, even though auroral lines can be more sensitive, their weaker intensities may hinder their utility. As we decrease $d$, the temperature in the proplyd outflow increases, so the ratio of auroral to nebular lines rises.

For example, we again consider the line ratios [\ion{O}{II}] 3726 \AA\ / [\ion{O}{III}] 5007 \AA\ and [\ion{O}{II}] 7320 \AA\ / [\ion{O}{III}] 5007 \AA. Here [\ion{O}{II}] 3726 \AA\ is the nebular line and [\ion{O}{II}] 7320 \AA\ the auroral line. Both ratios increase with $d$, as explained by the relative volumes of emission. As $d$ increases, the temperature in the flow decreases. This causes the emissivity of the auroral line [\ion{O}{II}] 7320 \AA\ to decrease more than the nebular line [\ion{O}{II}] 3726 \AA, see table \ref{tab:line_properties} for the excited state energy levels, hindering the growth of [\ion{O}{II}] 7320 \AA\ / [\ion{O}{III}] 5007 \AA\ compared to [\ion{O}{II}] 3726 \AA\ / [\ion{O}{III}] 5007 \AA. Both the density effect explained in section \ref{Density in the Outflow} and the temperature effect mean the line ratio [\ion{O}{II}] 3726 \AA\ / [\ion{O}{III}] 5007 \AA\ is far more sensitive to the FUV field.

\subsubsection{Properties of sensitive line ratios}

With an understanding of the physics that affects line ratios, we can now understand the general properties of CELs that produce the most sensitive line ratios. Suppose that we are considering a line ratio which increases as $d$ decreases. 
The line in the numerator should have a volume of emission that decreases, while the line in the denominator should have a volume that increases. Typically, this will include one line in ionization state II and one in ionization state III e.g. \ion{S}{II} and \ion{O}{III} (see Figure \ref{fig:I_fronts}). Closer to the ionizing source, the temperature in the outflow increases. The low ionization state line should come from a low energy state, so that the increased temperature for small $d$ has a limited contribution to its emissivity. However, the high ionization state line should come from a high energy state, so that the temperature rise contributes to an increase in the lines emissivity for small $d$. The critical densities should satisfy $n_\text{IF}>n_\text{crit,II}$ and $n_\text{IF}<n_\text{crit,III}$, where $n_\text{crit,II}$ and $n_\text{crit,III}$ are the critical densities of the excited state for the ionization state II and III species emission lines.
This means that the emissivity $j_\text{II}$ rises less than $j_\text{III}$ does for small $d$. An example from the MUSE wavelength range, which satisfies these conditions, is the line ratio [\ion{S}{II}] 6731 \AA\ / [\ion{O}{III}] 5007 \AA. We see that satisfying all three conditions results in a very strongly varying line ratio with $f=470$.

\subsection{Line ratio trends in stellar clusters}\label{stellar_clusters}

We now explore the extended parameter space by running a Monte Carlo simulation of line ratios in stellar clusters. The aim is to determine whether a spatial line ratio gradient should be observable across a stellar cluster. We do this for line ratios over a range of $f$ to determine an approximate threshold value for feasible observations.

Firstly, we consider a stellar cluster centred around a massive OB star, specified by its surface temperature and luminosity. In the first instance the star is $\Theta1$ Ori C. The stellar properties are in Table \ref{tab:177_341_properties}.

The stellar mass is sampled from the Kroupa initial mass function \citep{2001MNRAS.322..231K} given by 
\begin{equation}
    \xi(M_*) \ \ \propto \ \ 
    \begin{cases}
    \ \ \  M_*^{-1.3} & \text{for } \ \ 0.1 \ \text{M}_{\odot} \le M_* < 0.5 \ \text{M}_{\odot} \\[8pt]
    \ \ \  M_*^{-2.3} & \text{for } \ \ 0.5 \ \text{M}_{\odot} \le M_* \le 1 \ \text{M}_{\odot}.
    \end{cases}
\end{equation}
The lower limit and upper limits on stellar mass are chosen as 0.1 M$_\odot$ and 1 M$_\odot$, since this is the lower and upper limit in the FRIED grid. The radius of the disc $r_\text{d}$ is randomly chosen between $10$\,au and $50$\,au. The mass of the disc is randomly chosen between 10\% and 50\% of the maximum disc mass before gravitational instability,
\begin{equation}
    \frac{M_{\textrm{d,max}}}{M_*} = 0.17 \ \left( \frac{r_\text{d}}{100 \text{au}} \right)^{1/2} \ \left(  \frac{M_*}{M_{\odot}} \right)^{-1/2}
\end{equation}
\citep{2020MNRAS.494.4130H}.  Large discs are truncated and lose mass quickly, so a lower fraction of the maximum stable disc mass and smaller disc radii are deemed suitable.
The surface density at $1$\,au is constrained as
\begin{equation}
    \Sigma_{\text{au}} = \frac{M_\text{d}}{2 \pi  r_\text{d}\text{au}},
\end{equation}
assuming that the internal radius of the disc $r_\text{i} \ll r_\text{d}$. Finally, the distance from the UV source is randomly chosen so that the FUV is between $10^3 \ \text{G}_\text{0}$ and $10^6 \ \text{G}_\text{0}$ ($d=0.05 \ \text{--} \ 1.41$ pc). With these parameters, the FRIED mass loss rate can be determined. From a sample of N = 459 stars, 436 stars fell within the mass loss rate range $10^{-10} \ \text{M}_\odot/yr \ \le \ \dot{M} \ \le \ 10^{-6} \ \text{M}_\odot/yr$. Due to the small outside of this range they are neglected. Since in our model (for a given UV source) the system is defined by $d$ and $\dot{M}$, we produced a grid of models that span the ranges stated above. We interpolate the grid to find line ratios as a function of $d$ and $\dot{M}$.

\begin{figure*}
    \centering
    \includegraphics[width=\textwidth]{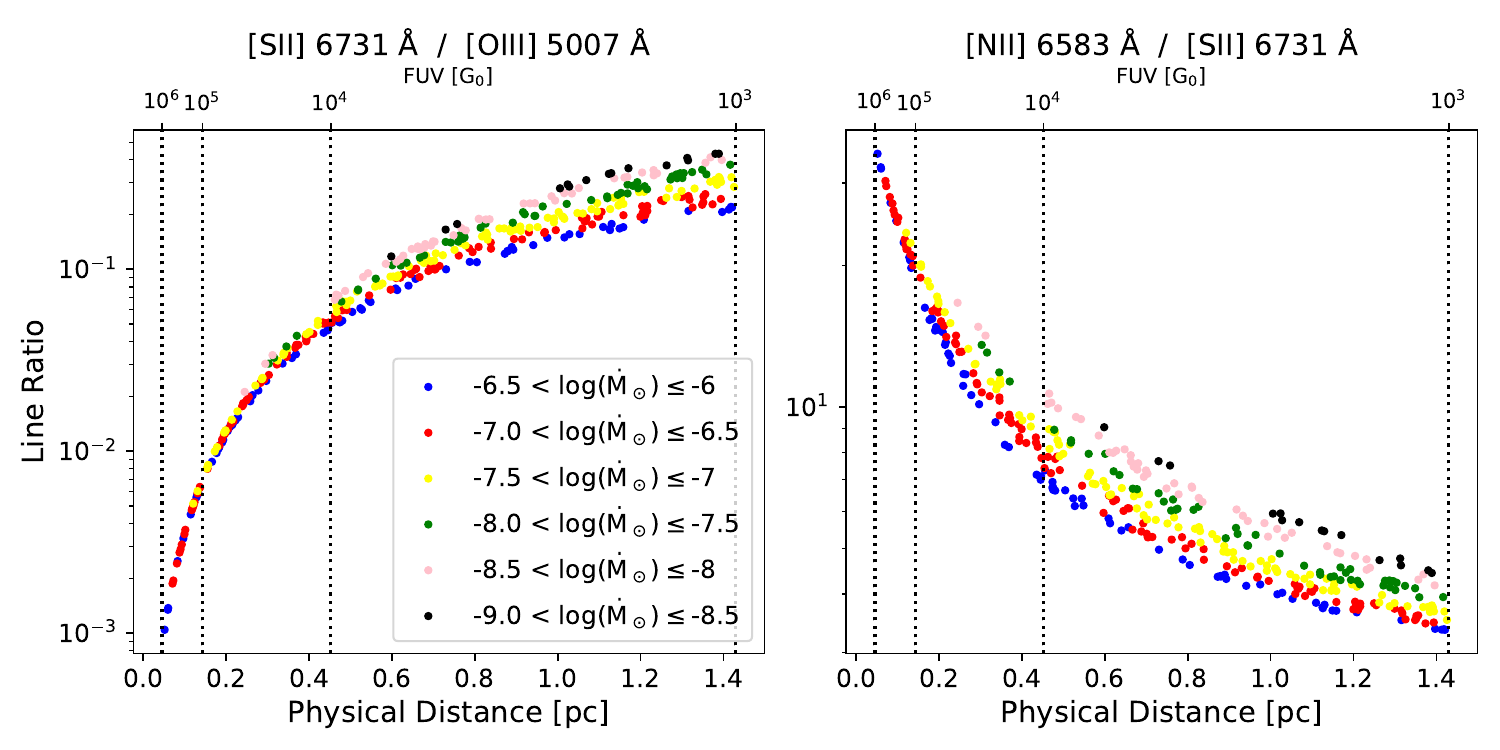} 
    \caption{Line ratios for our population of randomly sampled proplyds against their physical separation from the UV source. Vertical dotted lines show where the FUV field is $10^6$, $10^5$, $10^4$, and $10^3$ G$_\text{}0$ from left to right. The left hand panel shows the line ratio [\ion{S}{II}] 6731 \AA\ / [\ion{O}{III}] 5007 \AA\ and the right hand panel shows [\ion{N}{II}] 6583 \AA\ / [\ion{S}{II}] 6731 \AA. The mass loss rates span $10^{-9}$ -- $10^{-6}$ $\dot{\text{M}}_\odot$/yr, with the colours representing intervals within this full range. Spread in the clustering is due to a combination of the sampling selecting a wide range of mass loss rates at a given distance, and the line ratio being sensitive to mass loss rate at that distance.}
    \label{fig:MC_physical_distance}
\end{figure*}

In Figure \ref{fig:MC_physical_distance} for the sample of N = 444 stars the line ratios [\ion{S}{II}] 6731 \AA\ / [\ion{O}{III}] 5007 \AA\ and  [\ion{N}{II}] 6583 \AA\ / [\ion{S}{II}] 6731 \AA\ are shown against the physical separation $d$. In this case we assume that $d\tau(r)=0$ for all $r$. Both are strongly varying line ratios with $f = 470$ and $f=8.5$. The vertical black dotted lines from left to right represent FUV fields of $10^6,\ 10^5,\ 10^4,\ \text{and} \ 10^3$ G$_0$. The width of the cluster (the spread in line ratios at a given distance) depends on two factors i. the range of mass loss rates sampled at a given distance from the UV source, and ii. the sensitivity of the line ratio to those mass loss rates. The mass loss rate for each proplyd is indicated by colour in Figure \ref{fig:MC_physical_distance}. For example, for $G>10^5$ G$_0$ the line ratio [\ion{S}{II}] 6731 \AA\ / [\ion{O}{III}] 5007 \AA\ has a relatively narrow range in mass loss rates of $\approx 10^{-7.5}$ $\text{M}_\odot/\text{yr}$ -- $10^{-6}$ $\text{M}_\odot/\text{yr}$ and is insensitive to $\dot{M}$. Both contribute to a narrow cluster. For $G<10^4$ G$_0$ the range in mass loss rates grows to $\approx10^{-9}$ $\text{M}_\odot/\text{yr}$ -- $10^{-6}$ $\text{M}_\odot/\text{yr}$ and the line ratio becomes more sensitive to $\dot{M}$. Therefore, for greater $d$ the clustering grows. For line ratios sensitive to $d$ such as these, the line ratio is far more sensitive to $d$ than $\dot{M}$, so we see strong spatial line ratio gradients in stellar clusters. The line ratios are most sensitive at high $G$, so we expect that this technique will be most effective at identifying clusters in particularly high UV environments. The upper and lower edges of the clustering are due to the upper and lower limit on stellar masses of $1 \ \text{M}_\odot$ and $0.1 \ \text{M}_\odot$, respectively. There are few stars above $1 \ \text{M}_\odot$ in the IMF so we are not concerned with their effect on the clustering width. The Kroupa IMF describes an abundance of stars less than $0.1 \ \text{M}_\odot$. However, they would have very high mass loss rates and, therefore, would be short-lived. Therefore, we expect they would have little effect on the line ratio statistics. 

\subsubsection{Observational considerations}

To replicate realistic observations we need to account for the viewing angle of the proplyd, $i$, and the projected distance, $d_\perp$. To obtain these we assign each proplyd spherical coordinate angular positions $\theta$ and $\phi$. These are randomly chosen over the intervals (0 , $\pi$) and (0 , 2$\pi$), respectively. The viewing angle $i$ is given by
\begin{equation}
    i = \cos^{-1} \left( \sin\theta \sin\phi \ \right),
\end{equation}
where $i=0^{\circ}$ and $180^{\circ}$ correspond to the proplyd facing to and away from the observer, respectively. We produced a grid of $d$ and $\dot{M}$ for the inclinations $i=$ 0, 30, 60, 90, 120, 150, 180. For each proplyd, with equal probability, we assume the case that either any PDR or disc on the line of sight between the observer and the source completely obscures emission $d\tau(r<r_\text{IF}) \gg 1$ or that none is obscured $d\tau(r<r_\text{IF}) = 0$. This maximizes the potential noise introduced by absorption in the region $r_{\textrm{IF}}$, which preferentially obscures emission originating closer to $r<r_{\textrm{IF}}$ for more inclined proplyds, to ensure that we do not underestimate its impact. The projected distance can be calculated as 
\begin{equation}
    d_\perp = d \ ( \cos^2\theta + \sin^2\theta \ \cos^2\phi )^{1/2}.
\end{equation}
The complete set of randomly sampled parameters, the ranges they span, and their distributions are summarized in Table \ref{tab:free_parameters}.

{
\setlength{\tabcolsep}{7pt}
\renewcommand{\arraystretch}{1.5}
\begin{table}
    \centering
    \begin{tabular}{lcr}
    \hline
        Parameter & Distribution & Parameter Range \\
        Stellar Mass & Kroupa IMF & $0.1 \ \text{M}_{\odot} \le M_* \le 1 \ \text{M}_{\odot}$ \\
        Disc Radius & Uniform & $10 \ \text{au} \le r_\text{d} \le 50 \ \text{au}$ \\
        Disc Mass & Uniform & $ 0.1 \ M_{\textrm{d,max}} \le M_{\textrm{d}} \le 0.5 \ M_{\textrm{d,max}}$ \\
        FUV Field & Uniform & $10^3 \ \text{G}_\text{0} \le \text{FUV} \le 10^6 \ \text{G}_\text{0}$ \\
        Polar Angle & Uniform & $0 \le \theta \le \pi$ \\
        Azimuthal Angle & Uniform & $0 \le \phi \le 2\pi$ \\
    \hline
    \end{tabular}
    \caption{Summary of the free parameters required for our Monte Carlo analysis, the distribution which they are sampled from, and their range of allowed values.}
    \label{tab:free_parameters}
\end{table}
}

\begin{figure*}
    \centering
    \includegraphics[width=\textwidth]{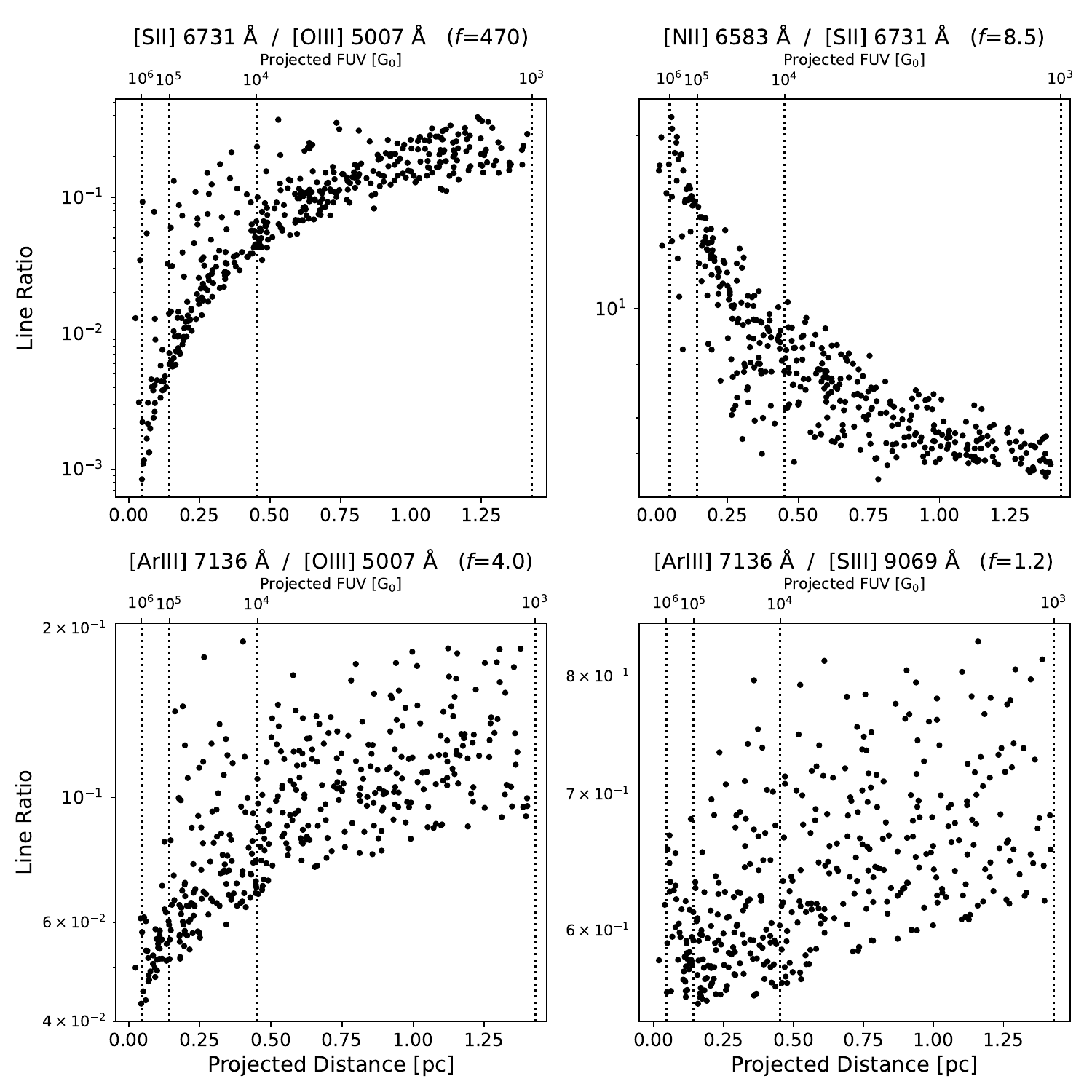} 
    \caption{Populations of line ratios which account for observational effects. We plot the line ratio against the projected distance and account for the proplyds angle of inclination. We consider the line ratios: [\ion{S}{II}] 6731 \AA\ / [\ion{O}{III}] 5007 \AA, [\ion{N}{II}] 6583 \AA\ / [\ion{S}{II}] 6731 \AA, [\ion{Ar}{III}] 7136 \AA\ / [\ion{O}{III}] 5007 \AA, and [\ion{Ar}{III}] 7136 \AA\ / [\ion{S}{III}] 9069 \AA, which have $f$ values of: 470, 8.5, 4, 1.2. These span a range so that we can see the value of $f$ which is sufficient for the trends to survive the noise. }
    \label{fig:MC_vary_grad_mean}
\end{figure*}

We now consider the same cluster of N = 459 proplyds and account for the observational effects. All points in Figures \ref{fig:MC_vary_grad_mean} and \ref{fig:MC_UV} have their own $(\phi,\theta)$ coordinate. Figure \ref{fig:MC_vary_grad_mean} shows the cluster in four different line ratios with values of $f=$ 470, 8.5, 4, and 1.2. These are [\ion{S}{II}] 6731 \AA\ / [\ion{O}{III}] 5007 \AA, [\ion{N}{II}] 6583 \AA\ / [\ion{S}{II}] 6731 \AA, [\ion{Ar}{III}] 7136 \AA\ / [\ion{O}{III}] 5007 \AA, and [\ion{Ar}{III}] 7136 \AA\ / [\ion{S}{III}] 9069 \AA. Both observational effects result in the clustering spreading. Since $d_\perp \le d$ for any $(\theta,\phi)$, plotting against the projected distance results in points moving left, that is, they appear to be at smaller distances than they really are. For our spherically symmetric cluster, this is by a mean value of $\bar{d}_\perp = \bar{d}/1.34$. The significance of the viewing angle depends on the line ratio under consideration. When the proplyd is facing the observer, all of its emission is observable. However, when the proplyd is facing away, a greater proportion of emission originating closer to the hydrogen ionization front is obscured by the PDR. Therefore, line ratios where the emission comes from different regions of the wind are most affected. Out of the line ratios in Figure \ref{fig:MC_vary_grad_mean}, the most affected line ratio is [\ion{S}{II}] 6731 \AA\ / [\ion{O}{III}] 5007 \AA. The viewing angle of the proplyd decreases the line ratio by at most a factor of 0.64 and a mean of 0.87. In this case [\ion{S}{II}] 6731 \AA\ emission originates closer to the ionization front than [\ion{O}{III}] 5007 \AA\ emission. Therefore, [\ion{S}{II}] 6731 \AA\ is preferentially obscured and the ratio decreases. For line ratios such as [\ion{N}{II}] 6583 \AA\ / [\ion{S}{II}] 6731 \AA, where each line is emitted from very similar regions, the effect is negligible. In general, the noise introduced by the projected distance is dominant over the viewing angle in reducing the clarity of the trends.

Weakly varying line ratios are more susceptible to the noise introduced by observational effects. Line ratios that vary by at least $\approx10$ times are necessary for robust identification of external photoevaporation in stellar clusters. These are highlighted in green in the sensitivity tables. For example, in Figure \ref{fig:MC_vary_grad_mean} the top two plots, [\ion{S}{II}] 6731 \AA \ / [\ion{O}{III}] 5007 \AA\ and [\ion{N}{II}]  6583 \AA\ / [\ion{S}{II}] 6731 \AA, provide sufficiently varying line ratios, whilst the bottom two, [\ion{Ar}{III}]  7136 \AA\ / [\ion{O}{III}] 5007 \AA\ and [\ion{Ar}{III}] 7136 \AA\ / [\ion{S}{III}] 9069 \AA, are deemed insufficiently varying.

\section{Discussion}\label{discussion}

\subsection{Dependence on the spectral type of the UV source}

\begin{figure*}
    \centering
    \includegraphics[width=\textwidth]{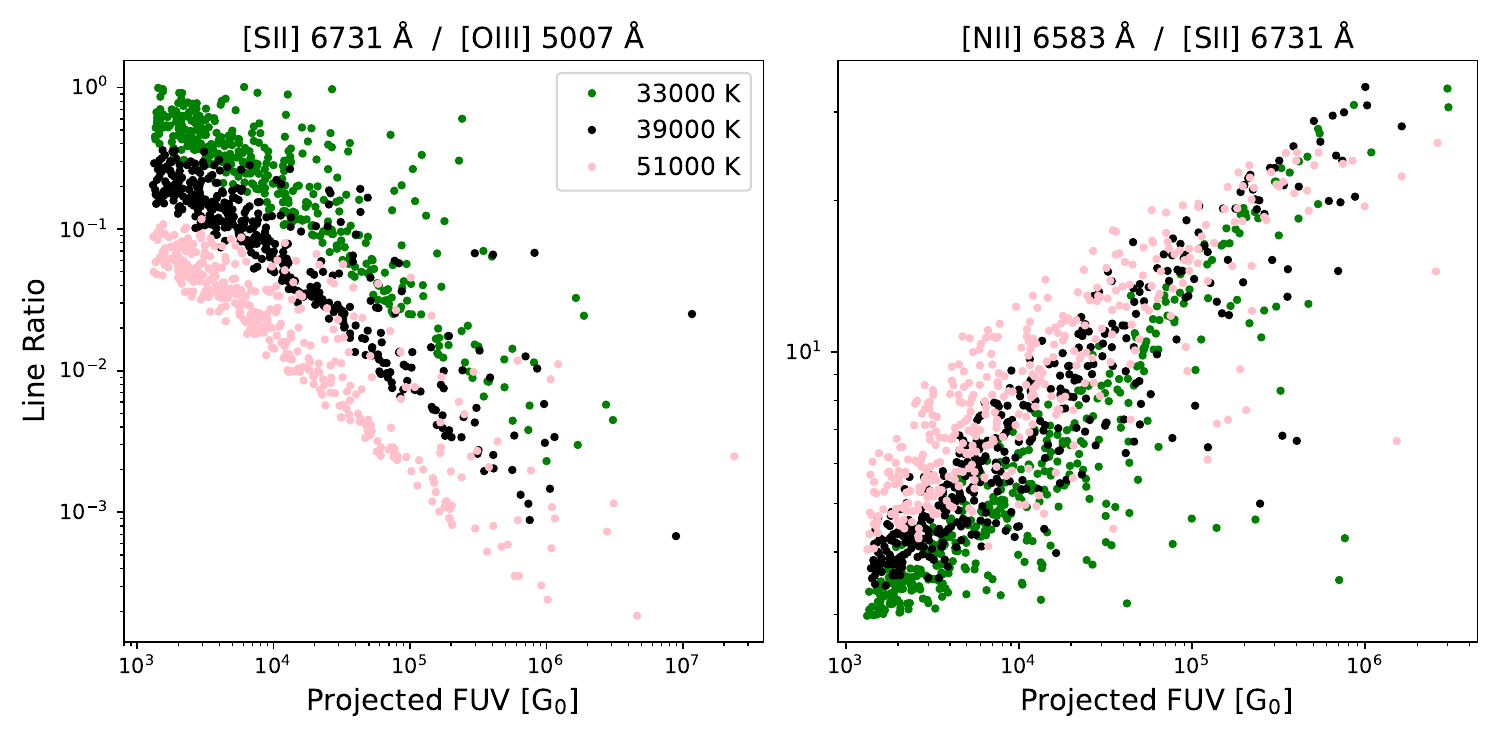} 
    \caption{Populations of line ratios, accounting for observational effects, for the stellar parameters shown in Table \ref{tab:MC_UV_stellar_prop}. We consider the strongly varying line ratios [\ion{S}{II}] 6731 \AA\ / [\ion{O}{III}] 5007 \AA\ and [\ion{N}{II}] 6583 \AA\ / [\ion{S}{II}] 6731 \AA. In this case we show the line ratio against the projected FUV field, which is the FUV field at the projected distance. The red is a 51,000 K star, blue 39,000 K, and the green is 33,000 K. The stellar surface temperature has little effect on the magnitude of the trends.}
    \label{fig:MC_UV}
\end{figure*}

For the strongly varying line ratios discussed previously, [\ion{S}{II}] 6731 \AA\ / [\ion{O}{III}] 5007 \AA\ and [\ion{N}{II}] 6583 \AA\ / [\ion{S}{II}] 6731 \AA, we consider the impact of the spectral type of the UV source. In Figure \ref{fig:MC_UV} we plot populations of line ratios for 3 different OB stars over an FUV field of $10^3$ G$_0$ to $10^6$ G$_0$, their stellar properties are shown in Table \ref{tab:MC_UV_stellar_prop}. The O7 star is $\Theta1$ Ori C and is the same 39,000 K star as used in section \ref{sec:results}. The remaining two are B0 and O3 stars with surface temperatures of 33,000 K and 51,000 K. Distances of approximately a few hundredths of a pc to a few pc from the massive star (specifics are in Table \ref{tab:MC_UV_stellar_prop}) correspond to FUV fields of between $10^3$ G$_0$ and $10^6$ G$_0$.

{
\setlength{\tabcolsep}{12pt}
\renewcommand{\arraystretch}{1.5}
\begin{table*}
    \centering
    \begin{tabular}{ccccc}
    \hline
        T$_*$ \ [k] & Stellar classification & L$_*$ \ [L$_\odot$] & $d$(FUV=$10^6$G$_0$) \ [pc] & $d$(FUV=$10^3$G$_0$) \ [pc] \\
        33,000 & B0 & $10^{4.9}$ & 0.023 & 0.93 \\
        39,000 & O7 & $10^{5.3}$ & 0.045 & 1.4 \\
        51,000 & O3 & $10^{6.0}$ & 0.085 & 2.7\\
    \hline
    \end{tabular}
    \caption{Summary of the parameters that we consider for alternative UV sources. We chose a range of stellar surface temperatures and luminosities consistent with OB stars \citep{2003ApJ...599.1333S}. In each case we randomly sample from distances where the FUV field is between $10^3$ G$_\text{0}$ and $10^6$ G$_\text{0}$. }
    \label{tab:MC_UV_stellar_prop}
\end{table*}
}

For both line ratios, the gradient of the trends is insensitive to the spectral energy distribution of the UV source. The value of the ratio for [\ion{S}{II}] 6731 \AA\ / [\ion{O}{III}] 5007 \AA\ shifts down for hotter stars. The greater fraction of EUV radiation decreases the relative volume of \ion{S}{II} emission, while increasing that of \ion{O}{III}. For [\ion{N}{II}] 6583 \AA\ / [\ion{S}{II}] 6731 \AA\ the variation is minimal because the relative volume of emission for both \ion{N}{II} and \ion{S}{II} decreases for hotter stars. The slight increase in ratio arises because, in terms of the hydrogen I-front, the \ion{S}{II}/\ion{S}{III} front is more sensitive to $d$ than the \ion{N}{II}/\ion{N}{III} front (see Figure \ref{fig:I_fronts}).

\subsection{Implications for future observations and observatories}
\label{sec:implications}
Our broad assessment of proplyd line ratio sensitivity to UV environment predicts some previously unused or underutilised diagnostics. These can be explored with existing and upcoming facilities, incorporating ultraviolet and infrared wavelengths.

For example, VLT/blueMUSE would operate in the range 3500 -- 5800 \AA\ \citep{2019arXiv190601657R}. From Table \ref{tab:line_ratios_ALL} that would enable us to utilise the predicted strong tracers of external photoevaporation in [\ion{O}{II}] 3726 \AA\ / [\ion{O}{III}] 5007 \AA\ and [\ion{O}{II}] 3726 \AA\ / [\ion{Ne}{III}] 3869 \AA, which are not available to VLT/MUSE. UV lines will be more affected by interstellar reddening. However, so long as the proplyds in a cluster experience similar extinction, this will have little impact on the overall trend. Other bluer options include UVEX which would operate in the range 1150 -- 2650\AA\ \citep{2021arXiv211115608K}. This would allow the use of the strongly varying line ratio [\ion{O}{II}] 3726 \AA\ / [\ion{C}{III}] 1909 \AA. Generally, in the UV, the line [\ion{O}{II}] 3726 \AA\ is a promising candidate line to form ratios with.

Infrared lines appear to have more limited use due to their generally lower luminosity, although it is up to the user to decide on a threshold luminosity. In the range 1,150 \AA\ -- 28.5 $\mu$m the left hand side of table \ref{tab:line_ratios_ALL} is populated largely by optical and UV lines. However, an exception is the line [\ion{S}{III}] 9531 \AA\ which is a particularly bright line and is sensitive in ratios with [\ion{N}{II}] 6583 \AA, [\ion{O}{II}] 3726 \AA, and [\ion{S}{II}] 6731 \AA\ which may make it particularly useful. Additionally, although they are not as bright as [\ion{S}{III}] 9531 \AA, the lines \ion{Fe}{II} 5.339 $\mu$m, \ion{Ne}{II} 12.81 $\mu$m, and \ion{Ar}{II} 6.983 $\mu$m still form generally sensitive line ratios. It is important to note that the lower utility of IR lines is in part a consequence of our model considering only emission from outside the I-front. In the infrared there are key tracers internal to the ionization front that we do not consider here and which have already demonstrated the tremendous utility of studying externally irradiated disks with JWST, particularly for astrochemistry \citep[e.g.][]{2023Natur.621...56B, 2024Sci...383..988B, 2023ApJ...958L..30R, 2024A&A...689L...4G}. Our models hence suggest that generally optical and UV lines are likely better used for unresolved identification of distant external photoevaporation. Follow up in the infrared with facilities such as the 30 m class ELT facilities will be able to resolve ionization fronts, determine mass loss rates, and study the composition \citep{2025OJAp....8E..54A}.

\subsection{Inside the ionization front}

Given the simple nature of our proplyd model, we have focused on emission lines that we expect to predominantly originate outside the proplyds ionization front. Here we reiterate that there are useful and important diagnostics of external irradiation interior to the ionization front. As mentioned in section \ref{sec:implications}, JWST has already illustrated its value for the study of the composition of externally irradiated disks  \citep[e.g.][]{2023Natur.621...56B, 2024Sci...383..988B, 2023ApJ...958L..30R, 2024A&A...689L...4G}. \cite{1998ApJ...502L..71S, 2023MNRAS.518.5563B} found that $G \gtrsim 5000$ G$_0$ the [\ion{O}{I}] 6300 \AA\ line luminosity rises significantly, and that the [\ion{O}{I}] to accretion luminosity ratio may provide a powerful diagnostic of external photoevaporation. The disk chemistry may also vary significantly with external radiation fields \citep{2013ApJ...766L..23W,2023ApJ...947....7B, 2023Natur.621...56B, 2025MNRAS.537..598K}, though the extent to which this is the case is still under study \citep{2023ApJ...958L..30R, 2024ApJ...969..165D}. Atomic carbon recombination lines have also been suggested as possible kinematic tracers of external photoevaporative winds, though detections have remained challenging so far \citep{2020MNRAS.492.5030H, 2022MNRAS.512.2594H}. More recently, \cite{2024A&A...692A.137A} found that the [\ion{C}{I}] 8727 \AA\ line, which is co-spatial with the disk, is significantly more prevalent in proplyds exposed to stronger FUV fields. Therefore, this may provide another efficient way to identify external photoevaporation. So, there is still a lot of known utility interior to the ionization front, as well as more work to be done to understand the emission in that zone.




\subsection{Caveats}

Any user of the line ratio trends presented in Table \ref{tab:line_ratios_MUSE} and \ref{tab:line_ratios_ALL} should bear in mind that in reality there will be additional sources of noise to what we applied in the stellar cluster simulations. Emission lines often have multiple physical origins of emission. Many of these come from the high velocity components in jets, knots, and bow shocks, e.g. \ion{Fe}{II}, \ion{S}{II}, and \ion{N}{II} emission in jets \citep{2015MNRAS.450..564R,2016MNRAS.463.4344R,2023A&A...670A.126F}, and \ion{Fe}{III}, \ion{S}{II}, and \ion{O}{II} emission in bow shocks \citep{2021MNRAS.502.1703M}. Therefore, the high-velocity component should be separated from the low-velocity component, which is dominated by the externally driven wind studied in our model. The tail of the proplyd itself is often a source of emission, e.g. \ion{S}{II} and \ion{N}{II} lines \citep{2024A&A...687A..93A}. Additionally, the gas phase abundances of some species can be especially variable, changing the luminosities of their emission lines. Massive stellar clusters often have multiple OB stars. For proplyds in between UV sources the spatial gradient of line ratios will likely look muddled. However, the outer proplyds should still show the same spatial trends in line ratios. The additional sources of noise we have discussed here will make the trends less clear. However, this can be compensated for by studying a sufficiently numerous stellar cluster.

Another important caveat is that line ratios that are close in wavelength offer the advantage that they are less susceptible to the effects of reddening, which is a factor that we have not considered in our evaluation of the line ratios. This can be accounted for, but we note that we are concerned with using line ratio variations to identify external photoevaporation across any given cluster where the reddening should be weakly varying \citep[e.g.][]{2010A&A...518A..62G, 2021ApJ...908...49F}.

\section{Summary and conclusions}\label{conclusion}

We have developed a model of the emission from the ionized wind of a proplyd i.e. external to the ionization front. The purpose of the model is to predict line ratios that are sensitive to the UV environment in order to provide the means of identifying external photoevaporation when proplyds are unresolvable. Since we are only interested in general trends, we opt for a simple model which can produce fast results, allowing us to study the parameter space of interest.

We assume that at the ionization front, where $T \approx 10^4$ K, the velocity is equal to the sound speed $v=c\approx10^6$ cm/s and is constant throughout the flow. Therefore, the density profile scales radially as $n\propto r^{-2}$, and the only free parameter in the wind is the mass loss rate. The mass loss rate is determined by the FRIED grid of mass loss rates which itself depends on the properties of the stellar system and the FUV field. We feed the density profile into \textsc{cloudy}, which returns to us the radial emissivity profile of any given emission line. Then, assuming that the emissivity profile is a function of radius only, defined over a hemisphere, we calculate the line ratio between any two lines. We compare the model to the proplyd 177-341W and find that the radial intensity profiles for nine well-observed lines are in good agreement for such a simple model, giving us confidence in the ability of the model to predict general trends. 
\vspace{\baselineskip}

Upon studying the behaviour of the model as a function of FUV field and proplyd mass loss rate our main conclusions are as follows:\\

\vspace{-\baselineskip}
\begin{enumerate}
    \item 
    We have developed a line-selection procedure in order to determine the emission line of each species most suitable for diagnosing external photoevaporation. For the base-case proplyd, the procedure considers both how sensitive its line ratios are with respect to the FUV field and how luminous the line is. Line ratios between all selected lines are displayed in sensitivity tables where the values, log$(f)$, represent the sensitivity of the line ratio. We propose that line ratios where $f \gtrsim 10$ have strong potential for identifying ongoing external photoevaporation. \\
    \item 
    The proplyd has three key properties which are sensitive to the FUV field strength and responsible for strongly varying line ratios. Firstly, the volume of the emitting species relative to the volume of the proplyd. Secondly, whether the critical density of a given emission line is above or below the density in the wind. Thirdly, the dependence of the line's emissivity on the temperature in the ionized wind. Consideration of these factors can aid in interpreting the cause for the line ratio trends in Tables \ref{tab:line_ratios_MUSE} and \ref{tab:line_ratios_ALL}. \\
    \item 
    For the most sensitive line ratios, $f\gtrsim 10$, the trends should be observable in stellar clusters. There are four key factors in our model which degrade the quality of the trend in a population: (1) the spread of mass loss rates at each distance (arising in the Monte Carlo sampling), (2) the sensitivity of the line ratio to mass loss rate at each distance (dependent on the properties of the wind), (3) observing the 2d projected distance rather than the physical 3d distance (a purely observational effect), and (4) the sensitivity of the line ratio to the proplyds angle of inclination (dependent on both the wind and the observations). However, consideration of other sources of line emission and noise must always be taken into account. \\
    \item
    The line ratio's sensitivity to the FUV field is insensitive to the spectral energy distribution of the UV source. Therefore, the same analysis can be applied to any cluster of stars without concern for the properties of the ionizing source. \\

    \item Purely from the perspective of identifying external photoevaporation through variation in line ratios, the best diagnostics are predicted to be in the UV and optical bands. Longer wavelength observations are still crucial for the detailed study of external photoevaporative disk winds of individual targets, for example, using ELT class facilities to resolve ionization fronts in regions like Carina in order to determine mass loss rates, and other longer wavelength instruments to study the disk/wind composition.

\end{enumerate}

\section*{Acknowledgements}
TP and TJH acknowledge funding from a Royal Society Dorothy Hodgkin Fellowship. TJH also acknowledges UKRI guaranteed funding for a Horizon Europe ERC consolidator grant (EP/Y024710/1)
MLA is funded by the European Union (ERC, WANDA, 101039452). 
WJH acknowledges financial support provided by  Dirección General de Asuntos del Personal Académico,
  Universidad Nacional Autónoma de México,
through grant   ``Programa de Apoyo a Proyectos de Investigación
  e Inovación Tecnológica IN109823''. We thank the reviewer, Uma Gorti, for their thorough review of our manuscript.

\section*{Data Availability}
Line ratio data for the base-case proplyd is available at \url{https://zenodo.org/records/15495918}.

The VLT/MUSE data that we compare with is publicly available (\url{https://telbib.eso.org/detail.php?bibcode=2024A%26A...687A..93A}). 

For the models, \textsc{cloudy} is a publicly available tool (\url{https://gitlab.nublado.org/cloudy/cloudy/-/wikis/home}), as is FRIED (\url{https://github.com/thaworth-qmul/FRIEDgrid}). Those combined with the density profile description here can reproduce the calculations in this paper.



\bibliographystyle{mnras}
\bibliography{example} 




\appendix

\section{Parker Wind}\label{A1}

We check that a Parker wind density profile gives line ratios with sensitivities similar to the constant velocity wind. Both start with a velocity, $v(r_{\text{IF}}) = c\approx 10^6$ cm/s, equal to the sound speed $c$ at the ionization front. The Parker wind model simultaneously solves the mass and momentum equation from the sonic point outward, i.e. $r \ge r_{\text{IF}}$. We use the following expression  (see e.g. \cite{1968Ap&SS...1..388D,1998AJ....116..322H})
\begin{equation}\label{Parker_wind}
    \frac{r}{r_{\text{IF}}}=\left[ \frac{c}{v(r)} \right]^{1/2} \ \text{exp} \left[   \frac{ v(r)^2  - c^2}{4c^2}  \right],
\end{equation}
derived by neglecting gravity in the momentum equation. The wind accelerates away from $r_\text{IF}$, resulting in a lower density profile in the wind than the constant velocity case. Therefore, in the Parker wind case the ionization front is at a smaller radius for the same mass loss rate. Since the final values in Table \ref{tab:line_ratios_MUSE} are calculated using simulations only at $10^3$ G$_\text{0}$ and $10^6$ G$_\text{0}$ (or $d=0.05$ pc and $d=1.41$ pc), it is quick to apply the Parker wind model to the same lines as in Table \ref{tab:line_ratios_MUSE} and verify the validity of the constant velocity model, see Table \ref{tab:line_ratios_MUSE_parker}.
{
\setlength{\tabcolsep}{2.5pt}
\renewcommand{\arraystretch}{1.5}
\begin{table*}
    \centering
    \begin{tabular}{c|ccccccccccccccccccccc}
        \hline
        species & \ion{O}{III} & \ion{H}{$\alpha$} & \ion{S}{III} & \ion{S}{II} & \ion{He}{I} & \ion{Ar}{III} & \ion{N}{II} & \ion{O}{II} & \ion{Cl}{III} & \ion{Fe}{III} & \ion{Fe}{II} & \ion{Ni}{II} & \ion{Si}{III} & \ion{C}{II} & \ion{Cr}{II} & \ion{Cl}{II} & \ion{Al}{II} & \ion{Mg}{II} & \ion{P}{II} & \ion{B}{II} \\ ($\lambda$ [\AA]) & \scriptsize{(5007)} & \scriptsize{(6563)} & \scriptsize{(9069)} & \scriptsize{(6731)} & \scriptsize{(7065)} & \scriptsize{(7136)} & \scriptsize{(6583)} & \scriptsize{(7320)} & \scriptsize{(8434)} & \scriptsize{(4986)} & \scriptsize{(7155)} & \scriptsize{(7378)} & \scriptsize{(8224)} & \scriptsize{(5113)} & \scriptsize{(8000)} & \scriptsize{(8579)} & \scriptsize{(7064)} & \scriptsize{(9218)} & \scriptsize{(7876)} & \scriptsize{(7030)} \\  \hline \ion{O}{III} \scriptsize{(5007)} & - & \cellcolor{c2} 0.89 & \cellcolor{c2} 0.57 & \cellcolor{c1} 2.75 & \cellcolor{c2} 0.34 & \cellcolor{c2} 0.66 & \cellcolor{c1} 1.86 & \cellcolor{c2} 0.57 & \cellcolor{c3} -0.28 & \cellcolor{c1} 2.94 & \cellcolor{c1} 1.9 & \cellcolor{c1} 1.75 & \cellcolor{c1} -2.1 & \cellcolor{c3} 0.3 & \cellcolor{c1} 1.9 & \cellcolor{c1} 1.49 & - & \cellcolor{c2} 0.42 & \cellcolor{c2} 0.96 & \cellcolor{c1} -1.55 \\ \ion{H}{$\alpha$} \scriptsize{(6563)} & \cellcolor{c2} -0.89 & - & \cellcolor{c2} -0.32 & \cellcolor{c1} 1.86 & \cellcolor{c2} -0.54 & \cellcolor{c3} -0.22 & \cellcolor{c2} 0.97 & \cellcolor{c2} -0.31 & \cellcolor{c1} -1.17 & \cellcolor{c1} 2.05 & \cellcolor{c1} 1.02 & \cellcolor{c2} 0.87 & \cellcolor{c1} -2.98 & \cellcolor{c2} -0.59 & \cellcolor{c1} 1.01 & \cellcolor{c2} 0.6 & \cellcolor{c1} -1.09 & \cellcolor{c2} -0.46 & \cellcolor{c3} 0.07 & \cellcolor{c1} -2.44 \\ \ion{S}{III} \scriptsize{(9069)} & \cellcolor{c2} -0.57 & \cellcolor{c2} 0.32 & - & \cellcolor{c1} 2.18 & - & \cellcolor{c3} 0.09 & \cellcolor{c1} 1.29 & - & \cellcolor{c2} -0.85 & \cellcolor{c1} 2.37 & \cellcolor{c1} 1.33 & \cellcolor{c1} 1.18 & \cellcolor{c1} -2.67 & \cellcolor{c3} -0.27 & \cellcolor{c1} 1.33 & \cellcolor{c2} 0.92 & \cellcolor{c2} -0.77 & - & \cellcolor{c2} 0.39 & \cellcolor{c1} -2.12 \\ \ion{S}{II} \scriptsize{(6731)} & \cellcolor{c1} -2.75 & \cellcolor{c1} -1.86 & \cellcolor{c1} -2.18 & - & \cellcolor{c1} -2.41 & \cellcolor{c1} -2.09 & \cellcolor{c2} -0.89 & \cellcolor{c1} -2.18 & \cellcolor{c1} -3.03 & \cellcolor{c3} 0.19 & \cellcolor{c2} -0.85 & \cellcolor{c2} -1.0 & \cellcolor{c1} -4.85 & \cellcolor{c1} -2.45 & - & \cellcolor{c1} -1.26 & \cellcolor{c1} -2.95 & \cellcolor{c1} -2.33 & \cellcolor{c1} -1.79 & \cellcolor{c1} -4.3 \\ \ion{He}{I} \scriptsize{(7065)} & \cellcolor{c2} -0.34 & \cellcolor{c2} 0.54 & - & \cellcolor{c1} 2.41 & - & \cellcolor{c2} 0.32 & \cellcolor{c1} 1.51 & - & \cellcolor{c2} -0.62 & \cellcolor{c1} 2.6 & \cellcolor{c1} 1.56 & \cellcolor{c1} 1.41 & \cellcolor{c1} -2.44 & - & \cellcolor{c1} 1.55 & \cellcolor{c1} 1.15 & \cellcolor{c2} -0.54 & - & \cellcolor{c2} 0.62 & \cellcolor{c1} -1.89 \\ \ion{Ar}{III} \scriptsize{(7136)} & \cellcolor{c2} -0.66 & \cellcolor{c3} 0.22 & \cellcolor{c3} -0.09 & \cellcolor{c1} 2.09 & \cellcolor{c2} -0.32 & - & \cellcolor{c1} 1.19 & \cellcolor{c3} -0.09 & \cellcolor{c2} -0.94 & \cellcolor{c1} 2.28 & \cellcolor{c1} 1.24 & \cellcolor{c1} 1.09 & \cellcolor{c1} -2.76 & \cellcolor{c2} -0.37 & \cellcolor{c1} 1.23 & \cellcolor{c2} 0.83 & \cellcolor{c2} -0.86 & \cellcolor{c3} -0.24 & \cellcolor{c3} 0.3 & \cellcolor{c1} -2.22 \\ \ion{N}{II} \scriptsize{(6583)} & \cellcolor{c1} -1.86 & \cellcolor{c2} -0.97 & \cellcolor{c1} -1.29 & \cellcolor{c2} 0.89 & \cellcolor{c1} -1.51 & \cellcolor{c1} -1.19 & - & \cellcolor{c1} -1.28 & \cellcolor{c1} -2.14 & \cellcolor{c1} 1.08 & \cellcolor{c3} 0.05 & \cellcolor{c3} -0.1 & \cellcolor{c1} -3.95 & \cellcolor{c1} -1.56 & \cellcolor{c3} 0.04 & - & \cellcolor{c1} -2.06 & \cellcolor{c1} -1.43 & \cellcolor{c2} -0.89 & \cellcolor{c1} -3.41 \\ \ion{O}{II} \scriptsize{(7320)} & \cellcolor{c2} -0.57 & \cellcolor{c2} 0.31 & - & \cellcolor{c1} 2.18 & - & \cellcolor{c3} 0.09 & \cellcolor{c1} 1.28 & - & \cellcolor{c2} -0.85 & \cellcolor{c1} 2.37 & \cellcolor{c1} 1.33 & \cellcolor{c1} 1.18 & \cellcolor{c1} -2.67 & \cellcolor{c3} -0.27 & \cellcolor{c1} 1.32 & \cellcolor{c2} 0.92 & \cellcolor{c2} -0.77 & - & \cellcolor{c2} 0.39 & \cellcolor{c1} -2.12 \\ \ion{Cl}{III} \scriptsize{(8434)} & \cellcolor{c3} 0.28 & \cellcolor{c1} 1.17 & \cellcolor{c2} 0.85 & \cellcolor{c1} 3.03 & \cellcolor{c2} 0.62 & \cellcolor{c2} 0.94 & \cellcolor{c1} 2.14 & \cellcolor{c2} 0.85 & - & \cellcolor{c1} 3.22 & \cellcolor{c1} 2.18 & \cellcolor{c1} 2.03 & \cellcolor{c1} -1.82 & \cellcolor{c2} 0.58 & \cellcolor{c1} 2.18 & \cellcolor{c1} 1.77 & \cellcolor{c3} 0.08 & \cellcolor{c2} 0.7 & \cellcolor{c1} 1.24 & \cellcolor{c1} -1.27 \\ \ion{Fe}{III} \scriptsize{(4986)} & \cellcolor{c1} -2.94 & \cellcolor{c1} -2.05 & \cellcolor{c1} -2.37 & \cellcolor{c3} -0.19 & \cellcolor{c1} -2.6 & \cellcolor{c1} -2.28 & \cellcolor{c1} -1.08 & \cellcolor{c1} -2.37 & \cellcolor{c1} -3.22 & - & \cellcolor{c1} -1.04 & \cellcolor{c1} -1.19 & \cellcolor{c1} -5.04 & \cellcolor{c1} -2.64 & \cellcolor{c1} -1.04 & \cellcolor{c1} -1.45 & \cellcolor{c1} -3.14 & \cellcolor{c1} -2.52 & \cellcolor{c1} -1.98 & \cellcolor{c1} -4.49 \\ \ion{Fe}{II} \scriptsize{(7155)} & \cellcolor{c1} -1.9 & \cellcolor{c1} -1.02 & \cellcolor{c1} -1.33 & \cellcolor{c2} 0.85 & \cellcolor{c1} -1.56 & \cellcolor{c1} -1.24 & \cellcolor{c3} -0.05 & \cellcolor{c1} -1.33 & \cellcolor{c1} -2.18 & \cellcolor{c1} 1.04 & - & \cellcolor{c3} -0.15 & \cellcolor{c1} -4.0 & \cellcolor{c1} -1.6 & \cellcolor{c3} -0.01 & \cellcolor{c2} -0.41 & \cellcolor{c1} -2.1 & \cellcolor{c1} -1.48 & \cellcolor{c2} -0.94 & \cellcolor{c1} -3.46 \\ \ion{Ni}{II} \scriptsize{(7378)} & \cellcolor{c1} -1.75 & \cellcolor{c2} -0.87 & \cellcolor{c1} -1.18 & \cellcolor{c2} 1.0 & \cellcolor{c1} -1.41 & \cellcolor{c1} -1.09 & \cellcolor{c3} 0.1 & \cellcolor{c1} -1.18 & \cellcolor{c1} -2.03 & \cellcolor{c1} 1.19 & \cellcolor{c3} 0.15 & - & \cellcolor{c1} -3.85 & \cellcolor{c1} -1.45 & \cellcolor{c3} 0.14 & \cellcolor{c3} -0.26 & \cellcolor{c1} -1.95 & \cellcolor{c1} -1.33 & \cellcolor{c2} -0.79 & \cellcolor{c1} -3.3 \\ \ion{Si}{III} \scriptsize{(8224)} & \cellcolor{c1} 2.1 & \cellcolor{c1} 2.98 & \cellcolor{c1} 2.67 & \cellcolor{c1} 4.85 & \cellcolor{c1} 2.44 & \cellcolor{c1} 2.76 & \cellcolor{c1} 3.95 & \cellcolor{c1} 2.67 & \cellcolor{c1} 1.82 & \cellcolor{c1} 5.04 & \cellcolor{c1} 4.0 & \cellcolor{c1} 3.85 & - & \cellcolor{c1} 2.4 & \cellcolor{c1} 3.99 & \cellcolor{c1} 3.59 & \cellcolor{c1} 1.9 & \cellcolor{c1} 2.52 & \cellcolor{c1} 3.06 & \cellcolor{c2} 0.55 \\ \ion{C}{II} \scriptsize{(5113)} & \cellcolor{c3} -0.3 & \cellcolor{c2} 0.59 & \cellcolor{c3} 0.27 & \cellcolor{c1} 2.45 & - & \cellcolor{c2} 0.37 & \cellcolor{c1} 1.56 & \cellcolor{c3} 0.27 & \cellcolor{c2} -0.58 & \cellcolor{c1} 2.64 & \cellcolor{c1} 1.6 & \cellcolor{c1} 1.45 & \cellcolor{c1} -2.4 & - & \cellcolor{c1} 1.6 & \cellcolor{c1} 1.19 & \cellcolor{c2} -0.5 & \cellcolor{c3} 0.12 & \cellcolor{c2} 0.66 & \cellcolor{c1} -1.85 \\ \ion{Cr}{II} \scriptsize{(8000)} & \cellcolor{c1} -1.9 & \cellcolor{c1} -1.01 & \cellcolor{c1} -1.33 & - & \cellcolor{c1} -1.55 & \cellcolor{c1} -1.23 & \cellcolor{c3} -0.04 & \cellcolor{c1} -1.32 & \cellcolor{c1} -2.18 & \cellcolor{c1} 1.04 & \cellcolor{c3} 0.01 & \cellcolor{c3} -0.14 & \cellcolor{c1} -3.99 & \cellcolor{c1} -1.6 & - & \cellcolor{c2} -0.41 & \cellcolor{c1} -2.1 & \cellcolor{c1} -1.47 & \cellcolor{c2} -0.94 & \cellcolor{c1} -3.45 \\ \ion{Cl}{II} \scriptsize{(8579)} & \cellcolor{c1} -1.49 & \cellcolor{c2} -0.6 & \cellcolor{c2} -0.92 & \cellcolor{c1} 1.26 & \cellcolor{c1} -1.15 & \cellcolor{c2} -0.83 & - & \cellcolor{c2} -0.92 & \cellcolor{c1} -1.77 & \cellcolor{c1} 1.45 & \cellcolor{c2} 0.41 & \cellcolor{c3} 0.26 & \cellcolor{c1} -3.59 & \cellcolor{c1} -1.19 & \cellcolor{c2} 0.41 & - & \cellcolor{c1} -1.69 & \cellcolor{c1} -1.07 & \cellcolor{c2} -0.53 & \cellcolor{c1} -3.04 \\ \ion{Al}{II} \scriptsize{(7064)} & - & \cellcolor{c1} 1.09 & \cellcolor{c2} 0.77 & \cellcolor{c1} 2.95 & \cellcolor{c2} 0.54 & \cellcolor{c2} 0.86 & \cellcolor{c1} 2.06 & \cellcolor{c2} 0.77 & \cellcolor{c3} -0.08 & \cellcolor{c1} 3.14 & \cellcolor{c1} 2.1 & \cellcolor{c1} 1.95 & \cellcolor{c1} -1.9 & \cellcolor{c2} 0.5 & \cellcolor{c1} 2.1 & \cellcolor{c1} 1.69 & - & \cellcolor{c2} 0.62 & \cellcolor{c1} 1.16 & \cellcolor{c1} -1.35 \\ \ion{Mg}{II} \scriptsize{(9218)} & \cellcolor{c2} -0.42 & \cellcolor{c2} 0.46 & - & \cellcolor{c1} 2.33 & - & \cellcolor{c3} 0.24 & \cellcolor{c1} 1.43 & - & \cellcolor{c2} -0.7 & \cellcolor{c1} 2.52 & \cellcolor{c1} 1.48 & \cellcolor{c1} 1.33 & \cellcolor{c1} -2.52 & \cellcolor{c3} -0.12 & \cellcolor{c1} 1.47 & \cellcolor{c1} 1.07 & \cellcolor{c2} -0.62 & - & \cellcolor{c2} 0.54 & \cellcolor{c1} -1.97 \\ \ion{P}{II} \scriptsize{(7876)} & \cellcolor{c2} -0.96 & \cellcolor{c3} -0.07 & \cellcolor{c2} -0.39 & \cellcolor{c1} 1.79 & \cellcolor{c2} -0.62 & \cellcolor{c3} -0.3 & \cellcolor{c2} 0.89 & \cellcolor{c2} -0.39 & \cellcolor{c1} -1.24 & \cellcolor{c1} 1.98 & \cellcolor{c2} 0.94 & \cellcolor{c2} 0.79 & \cellcolor{c1} -3.06 & \cellcolor{c2} -0.66 & \cellcolor{c2} 0.94 & \cellcolor{c2} 0.53 & \cellcolor{c1} -1.16 & \cellcolor{c2} -0.54 & - & \cellcolor{c1} -2.51 \\ \ion{B}{II} \scriptsize{(7030)} & \cellcolor{c1} 1.55 & \cellcolor{c1} 2.44 & \cellcolor{c1} 2.12 & \cellcolor{c1} 4.3 & \cellcolor{c1} 1.89 & \cellcolor{c1} 2.22 & \cellcolor{c1} 3.41 & \cellcolor{c1} 2.12 & \cellcolor{c1} 1.27 & \cellcolor{c1} 4.49 & \cellcolor{c1} 3.46 & \cellcolor{c1} 3.3 & \cellcolor{c2} -0.55 & \cellcolor{c1} 1.85 & \cellcolor{c1} 3.45 & \cellcolor{c1} 3.04 & \cellcolor{c1} 1.35 & \cellcolor{c1} 1.97 & \cellcolor{c1} 2.51 & - \\  \hline $\bar{l}/\bar{l}_{\text{H}\alpha}$ & 1.4 & 1.0 & 0.14 & 0.01 & 0.02 & 0.08 & 0.17 & 0.04 & $10^{-4}$ & $10^{-5}$ & $10^{-3}$ & $10^{-3}$ & $10^{-20}$ & $10^{-5}$ & $10^{-5}$ & $10^{-3}$ & $10^{-9}$ & $10^{-6}$ & $10^{-4}$ & $10^{-20}$ \\
        \hline
    \end{tabular}
    \caption{A Parker wind version of Table \ref{tab:line_ratios_MUSE} using the same lines. The values in the table represent the same quantity as in Table \ref{tab:line_ratios_MUSE}, but are calculated from a Parker wind model of the base-case proplyd at $10^3$ G$_\text{0}$ and $10^6$ G$_\text{0}$. The trends remain very similar, validating the use of the simplified constant velocity model.}
    \label{tab:line_ratios_MUSE_parker}
\end{table*}
}
The Parker wind model results in an ionization front which is $\approx 0.7$ times smaller than the constant velocity model. However, the ionization structure and trends in line ratios between Tables \ref{tab:line_ratios_MUSE} and \ref{tab:line_ratios_MUSE_parker} remain very similar. The line ratios most affected are those that are weakly varying, but these should not be used for diagnosing external photoevaporation anyway. Since the constant velocity and Parker wind are consistent with each other, we use the simpler constant velocity case for this work.


\bsp	
\label{lastpage}
\end{document}